\documentclass[10pt, a4paper]{article}
\usepackage[utf8]{inputenc}
\usepackage{cite}
\usepackage{amsmath,amssymb,amsfonts}
\usepackage{algorithmic}
\usepackage{algorithm}
\usepackage{graphicx}
\usepackage{textcomp}
\usepackage{booktabs}
\usepackage{subcaption}
\usepackage{xcolor}
\usepackage{geometry}
\title{Preference-Based Privacy Trading}

\newtheorem{theorem}{Theorem}
\newtheorem{lemma}{Lemma}

\newtheorem{corollary}{Corollary}
\newtheorem{definition}{Definition}

\author{
\uppercase{Ranjan Pal}
\uppercase{Yixuan Wang}
\uppercase{Swades De}
\uppercase{Bodhibrata Nag}
\uppercase{Pan Hui}
}

\date{}

\begin{document}

\maketitle

% -*- ispell-local-dictionary: "american"; TeX-master: "cloudshare.tex"; -*-
\begin{abstract}
In the modern era of the mobile apps (the era of \emph{surveillance capitalism} - as termed by Shoshana Zuboff) huge quantities of surveillance data about consumers and their activities offer a wave of opportunities for economic and societal value creation.
ln-app advertising - a multi-billion dollar industry, is an essential part of the current digital ecosystem driven by free mobile applications, where the ecosystem entities usually comprise consumer apps, their clients (consumers), ad-networks, and advertisers. Sensitive consumer information is often being sold downstream in this ecosystem without the knowledge of consumers, and in many cases to their annoyance. While this practice, in cases, may result in long-term benefits for the consumers, it can result in serious information privacy breaches of very significant impact (e.g., breach of genetic data) in the short term. The question we raise through this paper is: \emph{Is it economically feasible to trade consumer personal information with their formal consent (permission) and in return provide them incentives (monetary or otherwise)?}.
%However, the current personal data ecosystem is mostly de-regulated, fragmented, and inefficient. On one hand, end-users are often not able to control access (either technologically, by policy, or psychologically) to their personal data which results in issues related to privacy, personal data ownership, transparency, and value distribution. On the other hand, this puts the burden of managing and protecting user data on profit-driven apps and ad-driven entities (e.g., an ad-network) at a cost of trust and regulatory accountability. 
%Data holders (e.g., apps) may hence take commercial advantage of the individuals' inability to fully anticipate the potential (mis)uses of their private information, with detrimental effects for group privacy. 
%\emph{We address in this paper the problem of the existence and design of socially efficient consumer-data releasing ecosystems aimed at achieving a `group privacy-improving' welfare state amongst competing data holders.} 
In view of (a) the behavioral assumption that humans are `compromising' beings and have privacy preferences, (b) privacy as a good not having strict boundaries, and (c) the practical inevitability of inappropriate data leakage by data holders downstream in the data-release supply-chain, we propose a design of regulated efficient/bounded inefficient economic mechanisms for oligopoly data trading markets using a novel preference function bidding approach on a simplified sellers-broker market. Our methodology preserves the \emph{heterogeneous} privacy preservation constraints (at a grouped consumer, i.e., app, level) upto certain compromise levels, and at the same time satisfies information demand (via the broker) of agencies (e.g., advertising organizations) that collect client data for the purpose of targeted behavioral advertising. 
%Using concepts from supply-function economics, we characterize in theory, a rigorous privacy trading market design paradigm that always results in unique equilibrium (i.e, stable) market states that can be either economically efficient or inefficient, depending on the nature of market competition between data holders (app firms). Subsequently, we characterize in closed form, the efficiency gap (if any) at market equilibrium. 
%As a major finding, we show that increasing competition between app firms of similar market power for privacy trading activities contribute to increased societal privacy welfare, thereby suggesting regulators to enable privacy trading in segregated pools of similar app firms.
\end{abstract}

\section*{keywords}
information privacy, preference, supply function economics, trading, market equilibrium

\section{Introduction}
% \vspace{-2 pt}
\label{Section: Introduction}
Mobile applications (apps) are driving a major portion of the modern digital society, including business small and large as well as the state-of-the-art IoT/CPS systems. ln-app advertising is an essential part of this digital ecosystem of mostly free mobile applications, where the ecosystem entities comprise the consumers, consumer apps, ad-networks, advertisers, and retailers. As a popular example, \emph{Evite.com} may sell lists of their consumers attending a party in a given location to advertisers via ad-networks run by Google and Facebook. Similarly, the gene testing company \emph{23andMe} might sell their clientele information directly to pharmaceutical companies in order for the latter to develop medical drugs. As a social objective, a `win-win' deal between (a) the commercial interests of entities (e.g., enterprises, apps, databoxes) that aggregate and sell consumer data and those (e.g., ad-networks, retailers) that buy this data from the latter, (b) interests of consumer behavior targeting advertising firms, and (c) preserving consumer side information privacy (IP). The basic requirement for this `win-win' ecosystem to exist in the first place, is the flow of personalized information from the consumer to the advertisers and retailers via the ad-networks (or directly from consumer to the advertisers/retailers) for effective/profitable ad placements, that subsequently motivate the latter to collect personal data about consumers via apps. The vision and benefits for such an ecosystem were laid down by a certain school of information economists way back from the 70's (see more details in \cite{acquisti2016economics}), in favor of having increased aggregate societal welfare. More specifically, according to the survey, in return for personal data, advertisers and marketers will benefit the consumer side through monetary compensation (e.g., discounts, Facebook Libre coins) and intangible benefits (e.g., personalization and customization of information content), and price discrimination. Furthermore, the same school of information economists state that the lack of use of personal data might lead to opportunity costs and market inefficiencies. To furthermore emphasize the benefits of privacy trading, now from a consumer viewpoint, a survey conducted by the authors in \cite{benndorf2018willingness} advocate consumers willing to trade data for incentives. \emph{In this paper, we take the side of these economists to investigate privacy outcomes in society as a result of such markets.} However, before we lay down research contributions with respect to such markets, we provide an explanation of why such markets are a need of the day despite privacy concerns raised due to IP commercialization. 

\subsection{Need for Fair Privacy Commercialization}
Most would agree that doing business with consumer data without their consent is outright creepy. Consequently, as a landmark regulatory corrective step to prevent commercialization of personal data, the General Data Protection Regulation (GDPR) was initiated in May 2018 that impose constraints, rights, obligations, and voluntary consumer choice regarding  personal data and its use. 
However, it is questionable as to whether the psychological approach of many apps---in offering a binary voluntary opt in\slash out, often after presenting pages of legalese---results in user empowerment with respect to making the proper choice between gaining utility from an app versus not using it. Indeed, we see that individuals are increasingly using ad-blocking technology\footnote{https://pagefair.com/blog/2017/adblockreport/} as a means to `push-back', alongside deciding to gain utility from apps. However, ad blocking firms like \emph{Eyeo}, maker of the popular \emph{AdBlock Plus} product, has achieved such a position of leverage that it gets Google et.al., to pay it to have their ads whitelisted by default - under its self-styled `acceptable ads' program \cite{lomas_2019} - clearly going against the of the core functionality principle of ad-blockers. 

\noindent \emph{Thereby, with a significant likelihood, there might be an inevitable breach of personal consumer information in general to satisfy the economics behind the working of the current ad ecosystem}. 
According to a recent study \cite{lomas_2019} conducted post GDPR enactment, influential popular app-firms like New York Times (NYT) can likely make more revenues from traditional advertising channels such as TV/newspapers, compared to online/mobile advertising. However, this argument might not hold for moderate sized firms who consequently would rely heavily on behavioral advertising for generating revenues. The bottomline here is data intermediary entities will commercially gain from the consumer data release downstream, whereas psychologically tricked consumers, some of them being under the effect of the privacy paradox \cite{barnes2006privacy}, voluntarily give up their personal data and lose out on both privacy and monetary gains - \emph{an unfair proposition.} Moreover, one could argue here that paying for apps\footnote{There are quite a few services that already offer some level of choice/configuration between full subscription (no ads, thus no third party privacy exposure) and fully advertisement/analytics paid for (i.e. ``free"). Consequently
there's the possibility of doing an empirical study to populate
a model of peoples'(not yet evident that they are privacy-rational) ``willingness to pay" in terms of utility function/curves for privacy/money.} would mitigate this issue, however, statistics prove that consumers around the world are more keen on using free apps compared to paid apps\footnote{https://www.appsflyer.com/resources/state-app-spending-global-benchmarks-data-study/}, and are also quite neutral to the collection of cookies by third parties, during browsing activities\footnote{Statistic.com}. 

\noindent On an orthogonal (to regulatory issues) note, Shoshana Zuboff in her recent book \cite{zuboff2019age} states with numerous real-life surveillance examples of how since the early 2000's (primarily after 9/11), our daily life activities and `deepest secrets' are all recorded, rendered as behavioral data, processed, analysed, bought, bundled, and resold like sub-prime mortgages in a \emph{behavioral futures} market, thanks to companies such as Google and Facebook whose initial motivations for data collection were rooted in boosting ROI for their investors. 
%The litany of appropriated experiences is repeated so often and so extensively that we become numb, forgetting that this is not some dystopian imagining of the future, but the present. 
%Originally intent on organising all human knowledge, Google ended up controlling all access to it (the process starting post the 9/11 attacks when the US government became liberal on surveillance of human data for security purposes and also coinciding with Google needing to boost their ROI for their glamorous investors) ; we do the searching, and are searched in turn. Setting out merely to connect us, Facebook found itself in possession of our deepest secrets. 
And in seeking to survive commercially beyond their initial goals, these companies realised they were sitting on a new kind of asset: our `behavioural surplus', the totality of information about our every thought, word and deed, which could be traded for profit (via rejecting established norms of societal responsibility and accountability) in new markets based on predicting, shaping, and controlling our every need - or producing it. 
%In a move of such audacity that it bears comparison to the enclosure of the commons or colonial conquests, the tech giants unilaterally declared that these previously untapped resources were theirs for the taking, and brushed aside every objection. While insisting that their technology is too complex to be legislated, there are companies that have poured billions into lobbying against oversight, and while building empires on publicly funded data and the details of our private lives they have repeatedly rejected established norms of societal responsibility and accountability. What is crucially different about this new form of exploitation and exceptionalism is that beyond merely strip-mining our intimate inner lives, it seeks to shape, direct and control them. 
%Tech giant operations transpose the total control over production pioneered by industrial capitalism to every aspect of everyday life. 
The extraction of such information assets by tech giants is so grotesque, so creepy, that it is almost impossible to see how anyone who really thinks about it lives with it - and yet we do. There is something about its opacity, its insidiousness, that makes it hard to think about. Likewise the benefits of faster search results and turn-by-turn directions mask the deeper, destructive predations of what Shoshana Zuboff terms `surveillance capitalism', a force that is as profoundly undemocratic as it is exploitative, yet remains poorly understood - a central strategy of this regime. Despite more and more people expressing their unease about the surveillance economy, and seeking alternatives, it might be long before we extricate ourselves from the toxic products of both industrial and surveillance capitalism. \emph{Till then, one workable solution might be to trade consumer data with their consent in a fashion that benefits all fairly in the data release ecosystem, and not just the data greedy firms.} To this end, the reader is referred to our recently published work, \cite{rjcr}, for additional details on the rationale behind privacy trading being a solution jointly aligned with the supply and demand sides of a privacy market. 

%Finally, we comment on a mechanism that mitigates privacy issues from trading personal data.  

\subsection{Towards `Preference-Based' Trading }
A deeper look into existing research in the generic area of designing privacy preserving economic mechanisms (courtesy the survey paper in \cite{pai2013privacy}, though the paper is not in line with the idea of privacy trading as applicable to this work) reveals that the fundamental inability for \emph{any} economic mechanism dealing with consumer data to achieve a social optimal state with respect to privacy (be it for data trading ecosystems or otherwise) lie in (i) the hardness to satisfy \emph{strict} heterogeneous consumer privacy preferences, and (ii) the inability to internalize the negative externalities due to privacy leakage, e.g., recent Facebook-Cambridge Analytica data scandal \cite{facebook2018}. 
Thus, \emph{as our main idea, a direction towards optimizing social welfare, i.e., economic efficiency, is to relax the strictness of privacy preserving preferences}, thereby \emph{allowing heterogeneous consumers to compromise their ideal privacy requirements with their permission/consent in return for benefits} (e.g., monetary and non-monetary incentives). 
These benefits contribute to resolving the issue in (ii).

\noindent The weight behind this novel idea of ours lies in the fact that from a psychological perspective, most human beings are acceptable to making varied levels of compromises in real-life, especially for goods like privacy that have non-clear boundaries \cite{benndorf2018willingness} (See Section \ref{proof} for few examples where privacy compromises are acceptable). Note that privacy compromises by consumers would result in  apps selling more relevant personalized information to ad-networks (and thereby generating more revenue), the latter able to sell more ad-space to advertisers at an increased revenue, and the advertisers being able to target a broader personalized set of consumers. Thus, we have a win-win situation among all ecosystem entities. The big question then is: \emph{what is an optimal way to compromise aggregate consumer privacy?}  

\noindent \textbf{Research Goal} - As a major goal, we aim to investigate via a theory methodology, our radical idea of optimally compromising aggregate consumer privacy, in a simplified market ecosystem, through the combined use of micro-economic theory and a composition property characteristic of the family of information-theoretic privacy preserving technologies. Here, the term `optimal' is in the sense of achieving maximum utilitarian social welfare as an economic efficient state. Through our efforts, we wish to provide introductory foundational insights on designing information trading markets that improve social welfare, and pave the way for a more general analysis of complex trading markets. 

\subsection{Research Contributions}\label{Research_Contributions} 
We make the following research contributions in this paper.
\begin{itemize}
\item We model a privacy trading ecosystem setting as a supply-demand market consisting of (i) market competing (both, in perfect and oligopolistic fashion) data holders (DHs) representing app firms with locked-in consumer base and (ii) a single ad-network acting as a data broker between the app firms and the advertisers.   
A salient feature of this trading ecosystem is the use of data holder \emph{supply functions} \cite{klemperer1989supply} - privacy preference functions that map the amount of privacy compromise (the `supply') at an aggregate consumer level each data holder is willing to make, i.e., the supply, \emph{for} a given ``benefit" it receives from the ad-network per unit of data. The data holders submit their supply functions as bids to an ad-network that then executes a uniform market clearing ``benefit" mechanism for all competing data holders, to achieve optimal utilitarian privacy welfare at market equilibria (see Section \ref{System Model})\footnote{The readers are referred to the Section \ref{proof} (due to space constraints) for a qualitative introduction on supply function economics and its relevance to this work}.

\item We analyze perfectly competitive (in DHs) and oligopolistic privacy trading markets based on our proposed supply function model, for existence, uniqueness, and economic efficiency of market equilibria. For perfectly competitive markets we show that they achieve a maximum utilitarian social welfare state, i.e., an economic efficient state, at a unique equilibrium. However, for oligoplistic trading markets, we show that they reach a unique market equilibrium that does not maximize utilitarian privacy welfare in society (see Section \ref{Market_Analysis}). 

\item We mathematically characterize the efficiency loss for oligopolistic trading markets by quantifying the difference between the unique market equilibrium obtained in the competitive scenario with that in the oligopoly scenario, via a Price of Anarchy (PoA) measure. As major results, we find the following: (a) the set of data-holders at oligopolistic Nash equilibrium (ONE) who compromise on their privacy requirements at the aggregate consumer level, is a superset of that at the perfectly competitive equilibrium (PCE); (b) the market clearing ``benefit" (per unit of compromise) at the ONE is higher than that at the PCE, but the ratio of the two ``benefits" is bounded; (c) the sum total of data holder disutility (e.g., due to privacy compromise of their clients) at ONE is larger than that at PCE, but the ratio is bounded by certain mild assumptions; (d) if data holders have relatively homogeneous cost functions (e.g., for trading data types with similar privacy sensitivities), the differences between the PCE and ONE tend to be very small - if the cost functions are extremely heterogeneous (for trading data types with different privacy sensitivities), the quantification of the differences can serve as rules of thumb for the ad-network to limit the privacy compromising power of DH firms to promote utilitarian social welfare. For each of (a)-(d), we provide practical implications pertaining to privacy and policy. (see Section \ref{Efficiency_Analysis}). 

\item We show in Section \ref{Efficiency_Analysis} that for the problem at hand, our proposed supply function mechanism for privacy trading is \emph{optimal} over a feasible family of mechanisms. 
\end{itemize}

%\subsection{Evaluation Results}
 %For our experimental setting, we show in Figure 3 the results for benefit and supply function values at market equilibrium with respect to the number of iterations to market convergence. We observe that benefit and supply functions converge fast (within 25 iterations on a latest MacBook Pro with 16GB RAM) to the market equilibrium (ONE). This indicates the possibility of the existence of working markets satisfying all concerned stakeholders (as per our model) if personal data were to be traded. \emph{As part of future plans, we would like to run larger scale field experiments, conditioned on the availability of real data, to validate our speed and scalability claims on working privacy trading markets.} 

\section{Related Literature}
In this section, we briefly review related literature most relevant to privacy trading markets. We identified two strands of research in this context: one rooted in the economics literature, and the other rooted in the technical literature on privacy-aware mechanism design. With respect to privacy-preserving metrics of operation, applicable only to the technical literature, we note that the metric proposed in this work is assumed to fall in the same general family of metrics used in existing works, i.e., the family of information-theoretic privacy (IP) metrics (see \cite{wagner2018technical}) where resulting data is encapsulated with generated statistical noise to preserve IP, and IP guarantees are additive (e.g., as in differential privacy (DP)). 

\noindent The vision and benefits for information (privacy) trading (not necessarily consensual) had their roots in arguments made in the 1970s by University of Chicago economists, Posner\cite{posner1}\cite{posner2} and Stigler\cite{stigler}, in favor of having increased social welfare. In later years, their arguments were upvoted by information economists such as Laudon\cite{Laudon:1996:MP:234215.234476} and Acquisiti\cite{acquisti2016economics} Varian \cite{varian}, Odlyzko \cite{odlyzko}, Schwarz \cite{schwartz2003property}, and Samuelson \cite{samuelson2000privacy}. The primary thesis of these scholars being that the lack of use of personal client data will lead to opportunity costs and market inefficiencies (sub-optimal states of economic social welfare) since it conceals potentially relevant information from other economic agents (e.g., the downstream data intermediary entities in Figure 1) that eventually hamper the profitability of these agents. As a modern day example, client data (obtained via apps) on fitness, health habits, cyber-hygiene can benefit (cyber) insurance service agencies to target and allocate well-matched policies to their clients - conversely the lack of quality data can lead to bad matches and erode profit margins. In contrast to the Chicago-school views, a number of economists including Hirshleifer\cite{hirschleifer1}\cite{hirschleifer2}, Burke\cite{burke}, Wagman\cite{wagman},Daughety \& Reinganum\cite{daughety}, and Spence \cite{spence} are of the opinion that the costs to the demand side of the market to acquire quality client information in a non-consensual setting may outweigh its social benefit, thereby decreasing social welfare. It is here that consensual information trading with benefits to the supply side could reduce the costs to acquire supply side information and improve social welfare. \emph{In this work, we adopt the Chicago school of thought and assume that sellers will be consensual with the buyer demands in return for monetary remuneration.}

\noindent We assume consensual information trading to be regulated in the interest of social welfare, and an appropriate step for determining the effectiveness of trading in data intermediary settings such as in Figure 1. According to Varian\cite{varian}, Odlyzko\cite{odlyzko}, and Acquisiti\cite{acquisti2016economics}, consumer data obtained (with or without consent) can have negative effects on society simply because post transaction the consumers have little knowledge or control over how and by whom their personal data will later be used. The firm (e.g., ad-networks) may sell the consumer's data to third parties (e.g., advertisers), which may lead to spam and adverse price discrimination, among other concerns, and subsequently lead to consensual consumers opting out of trade in future. Regulation here can curb the adverse effects of these negative externalities arising from trading and significantly contribute to welfare efficient and complete markets (where supply equals demand) \cite{coase1960problem}\cite{contracttheory}. 
Examples of practical ways to implement regulations suggested in existing literature include legislative property rights on consumer personal data shared between the supply and demand side\cite{Laudon:1996:MP:234215.234476}, technical metrics (e.g., DP) being adopted by demand side data intermediaries (e.g., ad-networks) to check on the degree of IP breach\cite{rjcr}, and frameworks such as those developed in \cite{jain2019enhanced,madan2018privacy,thota2018centralized,al2019anonymous} to improve security and privacy for BigData systems (e.g., HDFS). 

\noindent Specifically, in relation to the data intermediary settings such as in Figure 1, De Corni`ere and Nijs\cite{corniere} rule out, for regulated consensual trading settings, direct price discrimination by the demand side on the supply side based on consumers' personal information by focusing instead on advertising firms' bidding strategies in auctions for more precise targeting of their advertisements. That is, given that consumers' private information provides a finer and finer segmentation of the population, firms can compete to advertise their non-discriminatory pricing over each of those consumer segments. Hence, by disclosing information about consumers, the ecosystem ensures that consumers will see the most relevant advertisements, whereas when no information is disclosed under a complete privacy regime, ads are displayed randomly. \emph{This is in contrast to our model that vouches for price-discrimination - the reason being in our setting, unlike the above-mentioned works, there is a statistical perturbation of the consumer private data sold downstream with noise for privacy considerations. Hence a finer clear segmentation is not possible.} 
De Corni et.al. also state that targeted advertising in the presence of private non-perturbed consumer information can lead to higher prices, and, in line with Levin and Milgrom\cite{levin}, Bergemann and Bonatti\cite{bergman}, and Cowan\cite{cowan} that improving match quality by disclosing consumer information to firms might be too costly to an intermediary - because of the informational rent that is passed on to selling firms. \emph{This is again in contrast to our findings - simply because in our model the selling data might be perturbed downstream by statistical noise.} 
%Research on privacy trading markets is scarce, being a fairly recent topic. In this section, we briefly review related literature most relevant to privacy trading markets - primarily intersecting the area of differential privacy and mechanism design, and propose differences in our approach inline where applicable. 
%With respect to metrics of operation, we emphasize that our work is applicable to general information gain privacy metrics as mentioned in \cite{wagner2018technical}, including differential privacy.

\noindent Most existing works on privacy-aware mechanism design \cite{ghosh2015selling} \cite{fleischer2012approximately} \cite{ligett2012take} \cite{roth2012conducting} \cite{ghosh2013privacy} \cite{nissim2014redrawing} \cite{ghosh2014buying} assume that there is a trusted data holder of unperturbed consumer data. The private data is either already kept by the data holder, noise perturbed by it, or is evoked using mechanisms that are designed with the aim of truthfulness. What the data holder purchases is the ``right" of using individuals' data in an announced way. 
\emph{A major direction in which our work differs from existing work is in considering that data holders are not trusted by consumers to keep their data private, may not noise perturb it to appropriate levels while releasing it to agencies like ad-networks, in return for benefits.} To this end, in the seminal work by \cite{ghosh2015selling}, individuals' data is already known to the data collector (the data collector here analogous to an ad-network in our work), and individuals (analogous to the data holder in our work) bid their costs of privacy loss caused by data usage, where each individual's  privacy cost is modeled as a linear function of $\epsilon$ if his data is used in an $\epsilon$-differentially private manner. The goal of the mechanism design here is to evoke truthful bids of individual cost functions. \emph{In contrast, our setting is more realistic and assume that (a) DH cost functions are private information - not for release to an ad-network, and (b) cost functions need not be linear but convex.} 

\noindent Subsequent works \cite{fleischer2012approximately}\cite{ligett2012take}\cite{roth2012conducting}\cite{nissim2014redrawing} explore various models for individuals' (analogous to DHs in our work) valuation of privacy, especially the correlation between the cost functions and the private bits. This line of work has been extended to the scenario that the data is not available yet and needs to be reported by the individuals to the data collector, but the data collector is still trusted \cite{ghosh2013privacy}\cite{xiao2013privacy}\cite{chen2016truthful}\cite{ghosh2014buying} - \emph{whereas we assume that the data collector (the ad-network in our case) is purposely selling consumer data (obtained via DHs) to advertisers for monetary gains.} For more details on the interplay between differential privacy and mechanism design, \cite{pai2013privacy} gives a comprehensive survey. In \cite{wang2016value}, the authors envisage a market model for private data analytics such that private data is treated as a commodity and traded in the market. In particular, the data collector (the ad-network in our case) uses a game-theoretic incentive mechanism to pay (or reward) individuals (DHs in our work) for reporting informative data, and individuals control their own data privacy by reporting noisy data with the appropriate level of privacy protection (or level of noise added) being strategically chosen to maximize their payoffs. \emph{However, unlike us, they assume that utility parameters of individuals are not private information, which may not be true in practice. In addition none of the above-mentioned works deal with the case of managing heterogeneous privacy guarantees across individuals (DHs in this work), as we do.} Very recently, the authors in \cite{khalili2019contract} address the heterogeneous privacy guarantee case. However, to address information asymmetry on the seller side, their solution is restricted to the design of a two-seller, single buyer contract based on a binary distribution of seller privacy attitudes. In contrast, our solution is general and addresses the multi-seller, single buyer setting, where seller preferences are captured using supply functions. 

\noindent In a very recent research effort, similar to our motivation, the authors in \cite{jin19} design a privacy trading mechanism for commercializing location privacy in mobile crowdsensing applications. More specifically, they propose an auction-theoretic framework between workers and the platform to trade location privacy data, given a differential privacy induced leakage budget. However, though they are similar in nature to our work in proportionalizing benefits with privacy leakage (and showing budget-balanced, truthful, and incentive compatibility properties of auction mechanisms), there are some significant differences between the contributions made in \cite{jin19} and this work: (i) we formally model market competition between established app firms serving a base of consumers; in contrast, the players (workers) in \cite{jin19} are mobile end users distributed in a geographical locality thereby only interacting with the platform through an auction, and not traditionally competing in an oligopoly market - hence such a market analysis is missing from their work, (ii) unlike us, the work in \cite{jin19} neither characterize market efficiency gaps in theory, nor do they prove the optimality of their mechanism over feasible families of economic variables (e.g., cost functions, mechanism classes, etc.), and (iii) as an obvious distinction, our application space, i.e., a supply-chain framework of mobile apps leaking data upstream to ad-networks and advertisers, is different in geographical scope from that of mobile crowdsensing.

\section{System Model}  \label{System Model}
In this section, we propose the salient features of our parameterized static market model representing a privacy trading ecosystem that is built atop the seminal economic theory of supply function bidding proposed by Klemperer et.al., in \cite{klemperer1989supply}, and \cite{johari2011parameterized}. Other applied works have built upon these seminal models \cite{pal2011settling, chen2010two,li2015demand,green1992competition,rudkevich1998modeling,baldick2001capacity,baldick2004theory}, and our efforts closely resemble that in \cite{li2015demand} (who also closely build their model atop \cite{johari2011parameterized}) due to the similarity in the demand-supply characteristics. 
Due to space constraints, we refer the reader to a qualitative background (see\cite{rjcr}) of supply function theory by Klemperer and Meyer as being an appropriate \emph{regulated} economic method that forms the primary basis in the design of markets to trade \emph{group privacy}\footnote{Shoshana Zuboff in her recent book, \emph{The Age of Surveillance Capitalism}\cite{zuboff2019age}, states that it is group privacy that is most important to surveillance capitalists as the individual user is just a pawn and not the product - the product is group data.} -  the privacy of a group of app clients, rather than individual clients themselves. Table \ref{tab:two} can be referred to for a set of important notations used in the paper.

% \vspace{-8 mm}
\begin{figure}[h]
\centering
 \includegraphics[scale = 0.24]{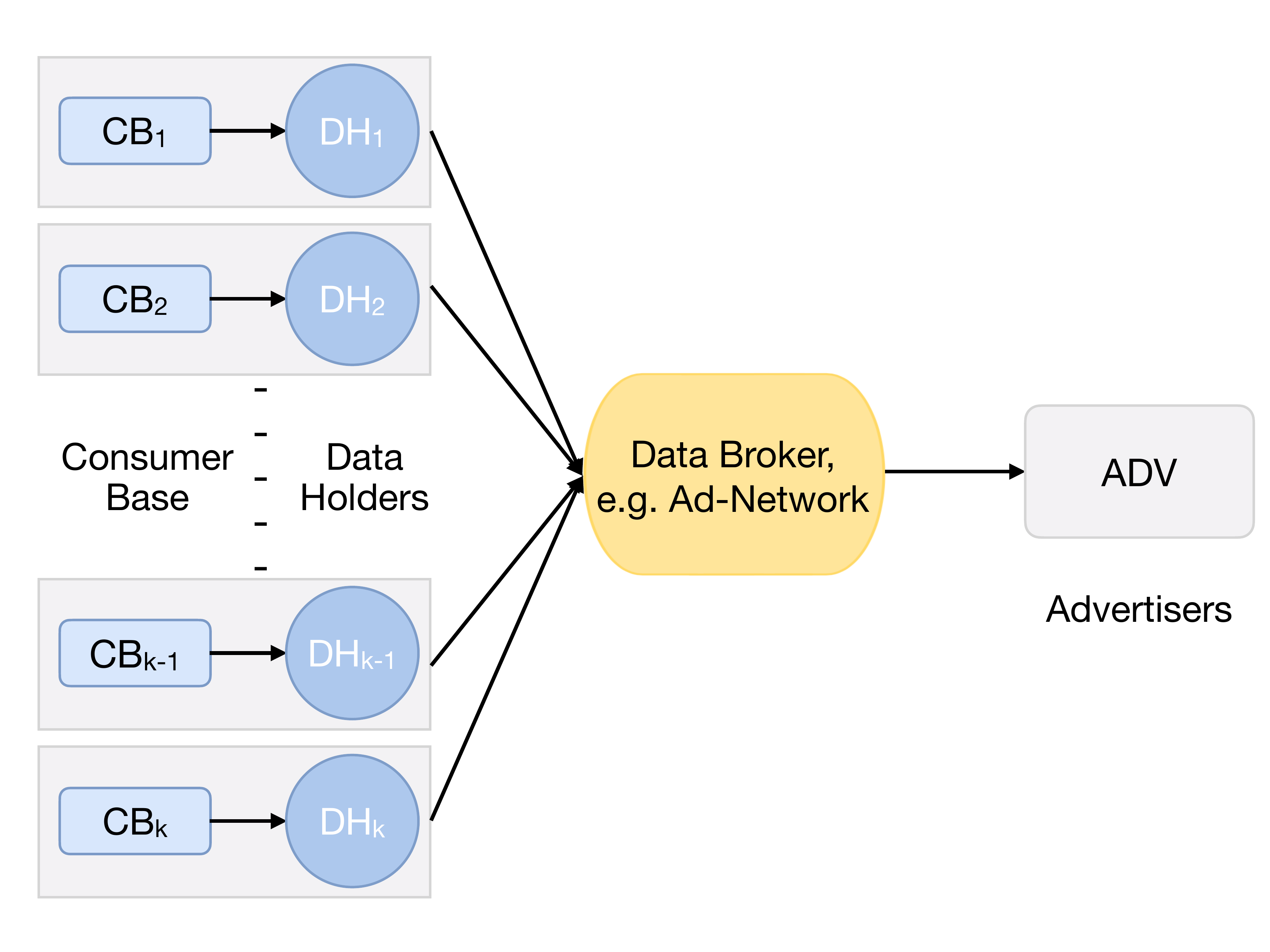} %architecture.pdf
  \label{fig:1}
%   \vspace*{-8 mm}
\caption{Market Architecture with a Single Data Broker (Ad-Network)}
\end{figure}
% \vspace{-6 mm}
\subsection{Market Elements}\label{Market_Elements}
Our market elements (see Figure 1) comprise of consumers locked in with their respective \textbf{data holders} (DHs) and an \textbf{ad-network} acting as a data broker between the data holders and a body of advertisers (ADV). We assume the presence of regulatory bodies (e.g., governments) whose goal is to ensure a certain level of social welfare state (e.g., maximum amount) keeping in mind the privacy interests of people in society. 

\noindent We assume that consumers are locked-in with their respective data holders in a given time period. Examples of data holders include ad-publishing mobile apps, social media apps, IoT databox apps\footnote{a given customer base can be associated with multiple competing app or social media DHs; however, in this work we assume a one-one mapping between consumers and DHs for relative tractable simplicity, as this setting itself is challenging enough. We leave the analysis of the one-many setting for future work.}, etc. Data holders compete with each other - as an example, competing mobile apps with similar functionalities (e.g., UberEats, GrubHub) are market competitors. Similarly, IoT databoxes manufactured by competing firms, each having their consumer base, compete with each other in the market. A consumer can simultaneously be client to multiple DHs. Based on pre-ordained policies, the data holders collect consumer data relevant to their functionality, and upon the consent of the consumers (e.g., Android and iOS phones have their own but different policies on how consumers can control data release to apps running on the phones). However, despite providing control to consumers, unwanted but voluntary data release by the latter is possible via methods designed through the proper use of psychology, behavioral economics, and neuroscience. Ad-networks (e.g., \emph{Google Ad Network}, \emph{Bing Ads} by Microsoft) act as mediators between DHs and advertisers, where the latter's goal is to post advertisements with DHs in order to enable targeting, tracking, and reporting of consumer impressions. Finally, to cite an example of the structure of data that could be traded by the DHs having access to aggregate consumer data from their client base - parts of it that is assumed to be private, a database is one of the possibilities. As popular practical examples, the firm \emph{BookYourData} (BYD) offers upstream buyers ready-made lists of contacts of business individuals across different industries, job titles, job functions, and job levels. A record in a list consists of contact information such as name, email, job function, department, country etc.
%\vspace{-0.69em} 
\subsection{Market Structure}\label{Market_Structure}
We consider two traditional market structures: \emph{perfect competition}, and \emph{oligopoly}, to be operative amongst the DHs. In each structure, the competing DHs trade privacy compromise amounts with a single ad-network\footnote{Since different ad-networks run their own supply function mechanisms for privacy trading independently of the others, the analysis of one extends to the others. Thus, each app will trade on different parameters with different ad-networks at market equilibrium (see Figure 3). Hence, in a somewhat simplistic sense, it is enough to analyse a single ad-network scenario. Moreover, when it comes to the number of major ad-networks, recent studies \cite{lomas_2019} report that they are primarily owned by \emph{Google} and \emph{Facebook}.} using a supply function bidding\footnote{Supply side privacy preferences, as functions of incentives, derived via survey Q\&A, deviates us from the use of the standard Bertrand and Cournot trading mechanisms that have one-dimensional (price or quantity) strategy spaces.} approach (see Section \ref{System Model}.C). The ad-network in return provides some ``benefits" (to be explained later in this section) to the DHs based on the amount of compromise made by the DHs. The ADV generates a demand\footnote{This is usually done through a bidding process like Vickrey-Clarke-Groves (VCG) auction (not the explicit focus of this work) between the ADVs and the ad-network, based on consumer data that interests relevant ADVs.} for consumer information to the ad-network, and in pay the ad-network to match them with appropriate DHs so as to enable targeting, tracking, and reporting of consumer impressions.
%\subsection{Structure of Data to be Traded}
%To cite an example of the structure of data that could be traded by sellers (e.g., mobile apps) having access to aggregate consumer data from their client base, parts of it that is assumed to be private, we choose a database to be a likely structure. As popular practical examples, the firm \emph{BookYourData} (BYD) offers upstream buyers ready-made lists of contacts of business individuals across different industries, job titles, job functions, and job levels. A record in a list consists of contact information such as name, email address, job function, job department, country etc. The organization \emph{SalesLead} (SL) maintains a variety of datasets of American businesses in the form of profession-based lists and state/province-based lists - the \emph{Accountant Sales Leads} dataset contains records of US-based accountants, whereas the \emph{Alabama Sales Leads} dataset contains records of different businesses (accountants, real-estate agents, etc.) based in Alabama. Each record in a dataset consists of contact information such as mailing address, geo-location, email address, phone number, etc. As another major example, the telemarketing company \emph{TelephoneLists} specializes in offering its buyers phone lists as datasets that consists of information on consumers (contact details, demographics, etc.) as well as businesses (number of employees, sales, etc.) in North America.
% \vspace{-1 mm}

\subsection{Model for Supply Function Bidding}
\label{Parameterized}
In this section we formally introduce the mechanism between competing DHs and the ad-network. A diagrammatic illustration of the process as shown in Figure 2. 
% \vspace{-5 mm}
\begin{center}
\begin{figure}[h]
\centering
 \includegraphics[width=0.8\linewidth]{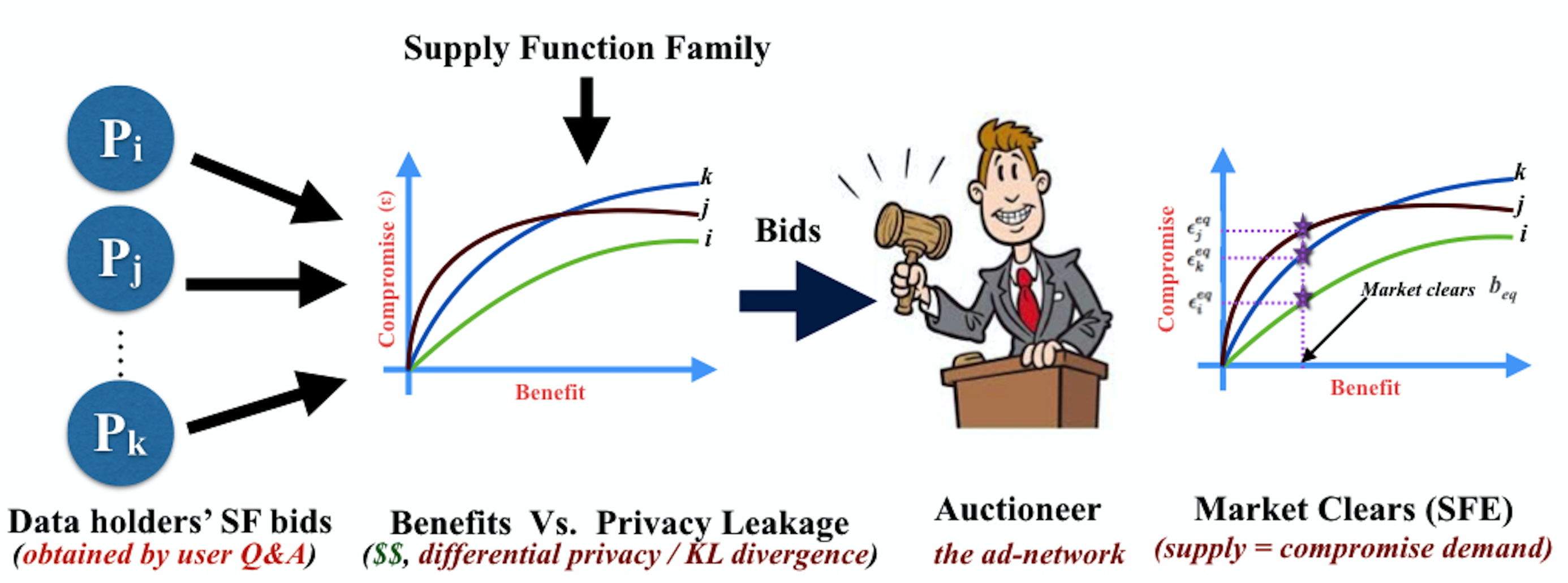}
\caption{Illustrating Privacy Preference Function Trading with One Broker}
 \label{fig:2}
\end{figure}
\end{center}
% \vspace{-10 mm}
\begin{table}[h]%
\caption{Table of Important Notations}
\label{tab:two}
\begin{minipage}{\columnwidth}
\begin{center}
\begin{tabular}{ll}
  \toprule
  $N$, $n = |N|$ & set and number of data holders, i.e., DHs\\
  $q_i$ & privacy compromised amount for DH $i$\\
  $p_i$ & per unit of compromise benefit of DH $i$\\
  $b$ & bidding parameter\\
  $b^{\ast}$ & Nash equilibrium bidding profile\\
  $C_i$ & cost function for DH $i$\\
  $u_{i}$ & utility function of DH $i$\\
  $d$ &  privacy compromise threshold\\
  $ONE$ &  oligopolistic Nash equilibrium\\
  $PCE$ &  perfectly competitive equilibrium\\
  $S_i$ & privacy compromise amount, DH $i$ willing to take\\
  $\pi_i$ &  payoff for DH $i$\\
  \bottomrule
\end{tabular}
\end{center}
\bigskip\centering
\end{minipage}
\end{table}
%\vspace{-0.55em} 
% \vspace{-8 mm}
\noindent \textbf{Setup -} Consider a set $N$ of $|N|$ DHs that are locked-in with their respective consumer base. In the ideal state, each DH needs to obey certain privacy requirements derived from the privacy preferences of their consumer base. To preserve generality, we assume that the privacy requirements of each DH map to a privacy metric that is an element of the set of \emph{information gain metrics} \cite{wagner2018technical} that measure the amount of information an adversary can gain. Note that the differential privacy metric is just one element of this set. Higher the value of the privacy metric, the less information an adversary can gain. However, given the presence of the ad-network and ADVs, \emph{there are two main reasons why there may not be the simultaneous satisfaction of privacy requirements of each DH}: (i) keeping in mind the ``benefit" making mindset of DHs (the ``benefit" whose source are the ADVs), achieving the optimal cost-benefit tradeoff with the ad-network might not guarantee strict privacy-preservation for DHs, (ii) it is known, via results from \cite{pai2013privacy}, that designing mechanisms that ensure heterogeneous privacy preservation at a utilitarian social welfare optimal state is an open problem. 

\noindent \textbf{The Process - } Each DH $i \in N$ is willing to consensually compromise $q_{i}(b_{i}, p_{i})$ amounts of aggregate client privacy (measured through the privacy metric - as shown in Figure 2, usually either DP, KL-divergence, Mutual Information, etc.) with the ad-network, in return for a per-unit of compromise benefit value, $p_{i}$, i.e., $q_{i}$ is a parameterized function of $p_{i}$ and a non-negative bidding parameter $b_{i}$. As an example, let $q_{i}$ to be a linear function of the form:
\begin{equation}
q_{i}(b_{i}, p_{i}) = b_{i}p_{i},\, i\in N,
\end{equation}
The compromise function, $q_{i}$, for each DH $i$ is their \emph{parameterized supply function}. The benefit to each DH, $p_{i}$ from the ad-network is primarily monetary in nature. Examples of benefits to the consumer base (derived from $p_{i}$\footnote{DHs make up for the discounts through benefits from the ad-network.}) include the amount of \emph{price reduction} over the market price paid by individual consumers locked-in with a given DH\footnote{The consumer market prices charged by competing DHs might vary for each DH.} (for the case of paid apps), or in the case DHs are free to consumers - an amount of \emph{reduction in the number of advertisements} displayed on the DH at a time instant (e.g., in case of an app) for each consumer to improve their usability experience.  

\noindent We emphasize here that each DH $i$ only submits the function $q_{i}$ to the ad-network, as a signal of its preference on privacy compromise, without revealing its private utility/payoff function (see Section \ref{Market_Analysis}) of which $q_{i}$ is just a part. Subsequently, the ad-network just has the values of $q_{i}$'s at its disposal to arrive at a market uniform market clearing value of per-unit benefit that maximizes social welfare amongst the DHs\footnote{One could argue that the popular Kelly's mechanism would also suffice to obtain social welfare optimality, but the latter mechanism is suitable only for one-dimensional bids, and not necessarily functions.} 

\noindent We assume that the total privacy compromise demand for the ad-network coming upstream from the advertisers end needs to meet a specific amount $d > 0$ (for a general information-theoretic privacy measure) \footnote{In the special case when the privacy metric under consideration is differential privacy, the total compromise demand $d$ is analogous to the quantity $\epsilon_{d}$ from Section \ref{System Model}, where $\epsilon_{d}$ = 0 denotes a situation of zero compromise.} when it clears the market, i.e.,
\begin{equation}
\sum_{i}q_{i}(b_{i}, p) = \sum_{i}b_{i}p = d,
\end{equation}
or 
\begin{equation}
p(b) = \frac{d}{\sum_{i}b_{i}}.
\end{equation}
Note here that Equation (2) holds due to the composability property of certain privacy metrics such as differential privacy \cite{dwork2006differential}\cite{dwork2014algorithmic}.  
$\boldmath{b} = (b_{1}, ....,b_{N})$ is the supply function profile of the DHs. In the event when $\sum_{i}b_{i} = 0$, the ad-network will reject the bid.

 % Table

\section{Markets Analyses}\label{Market_Analysis}
In this section, we analyze perfectly competitive and oligopolistic market structures of DH competition in the backdrop of a single ad-network. The strategy space for the DHs is the set of feasible parameter values for their supply functions. We assume no restrictions on DH compromise amounts and select the linear supply function as the preferred choice for the DHs. To this end, we first provide a strong rationale on our choice of supply function. We then proceed with the markets analyses in view of the development in \cite{li2015demand}. 

\noindent \textbf{Why Use a Linear Supply Function?} - We answer this question by first stating that, unlike us, the seminal work in \cite{klemperer1989supply} does use a general function as the bidding strategy for the purpose of analysis. However, if our bidding action were to change from the linear form (represented by the single variable, $b_{i}$ in our work) to a general form like in \cite{klemperer1989supply}, the analysis of the strategic behavior of the DHs become much more complicated. To drive home this point, solving the general supply function equilibrium (SFE) (introduced in \cite{klemperer1989supply}) requires solving a set of differential equations. To the best of our knowledge, there are only \textit{existence} results about the SFE while assuming the agents (DHs in our work) are symmetric (i.e., with the same cost function) or assuming there are only two asymmetric agents - \emph{these assumptions are not practical in reality.}  
For practical applications, the asymmetric case is more interesting. On the positive side, the greatest advantage of using linear supply function over the general forms is the ability to handle asymmetric DHs when there are more than two DHs. Moreover, as we will show later in this section, (a) the linear supply function allows us to get a closed form characterization for the structure and efficiency of the market equilibria, which could be impossible to get if using the general supply function, and (b) \emph{in the case of oligopoly markets, linear supply function induced markets minimize worst case efficiency loss for non-restricted compromise markets.} Thus, we lose no generality in working with linear supply functions as they would be incentive compatible for rational DHs to use (see Section \ref{Efficiency_Analysis}). 

\subsection{Perfectly Competitive Markets}
In perfectly competitive markets, DHs are `benefit taking'. Such markets arise when there are a plethora of DHs selling similar basic consumer information (e.g., users' preferences towards the items or products, language preference, time zone) that are mostly not very personal - so a standard common benefit value ensues. Given a benefit $p$, each DH $i$ maximizes its net revenue given as:
\begin{equation}
\max_{b_{i} \ge 0}pq_{i}(b_{i}, p) - C_{i}(q_{i}(b_{i}, p))
\end{equation}
where the first term is the revenue of DH $i$ when it compromises $q_{i}(b_{i}, p)$ amount of privacy at a benefit $p$ per unit of compromise with a bidding parameter of $b_{i}$, and the second term is the total cost incurred to make the compromise. This cost can be interpreted as the sum of (a) the cost of making technical adjustments required to compromise privacy (e.g., technological/software costs of hosting ads by advertisers), (b) costs of handling consumer complaints/unpopularity, (c) brand/app switching with respect to degradation of quality of experience (QoE) arising from clients experiencing delay and high cellular bandwidth costs in loading apps.

\begin{definition}
\emph{\noindent A perfectly competitive equilibrium (PCE) for the privacy compromise system is defined as a tuple $\{(\bar{b_{i}})_{i\in N}, \bar{p}\}$ such that $\bar{p_{i}}$ is optimal in (4) for each DH $i$ given the benefit $\bar{p}$ and $\sum_{i}q_{i}(\bar{b_{i}}, \bar{p}) = d$. }
\label{definition1}
\end{definition}

\noindent The following result shows the existence and uniqueness of PCE, and it also shows the efficiency of the latter in maximizing utilitarian social welfare. The proof of the theorem is in the Section \ref{proof}. 

\begin{theorem}
\emph{The PCE, $\{(\bar{b_{i}})_{i\in N}, \bar{p}\}$, for the privacy compromise system exists and is efficient, i.e., $(\bar{q}_{i})_{i\in N} = (q_{i}(\bar{b}_{i}, \bar{p}_{i}))_{i\in N}$ maximizes the utilitarian social welfare amongst the DHs expressed mathematically as follows:
$max_{q_{i} \ge 0}\sum_{i} - C_{i}(q_{i})$, subject to $\sum_{i}q_{i} = d$.
If the cost function $C_{i}(q_{i})$ is strictly convex, the PCE is unique. }
\label{theorem1}
\end{theorem}

\noindent \textbf{Theorem Implications} - The theorem implies that there exists a pure (and unique, if DH cost functions are strictly convex) strategy PCE vector of DH privacy compromise amounts for all DHs at a particular homogeneous PCE benefit $\bar{p}$ set by the ad-network that meets the aggregate ad-network demand of $d$ units of total privacy compromise, and maximizes utilitarian social welfare amongst the DHs. \emph{In a nutshell, the theorem states that at market equilibrium efficient privacy trading is possible amongst heterogeneous DHs and an ad-network.}

\noindent Based on the above theorem, we can further study how the compromise cost function affects a DH's privacy compromise amount at PCE. For each DH $i$, we define the base privacy compromise marginal cost as $C^{0}_{i} = C^{'}_{i}(0^{+})$. Without loss of generality, we assume that  $C^{0}_{1} \le C^{0}_{2} \le .......\le  C^{0}_{|N|}$. For modeling convenience, we also introduce parameter $C^{0}_{|N| + 1}$ and set its value to $C^{'}_{n}(d)$. Thus, we have  $C^{0}_{1} \le C^{0}_{2} \le .......\le  C^{0}_{|N|} \le C^{0}_{|N| + 1}$. We have the following result on the privacy compromise characteristics of individual DHs, the proof of which is in the Section \ref{proof}. 

\begin{theorem}
\emph{Let $\{(\bar{b}_{i})_{i\in N}, \bar{p}\}$ be a PCE and $\bar{q}_{i}$ = $q_{i}(\bar{b}_{i}, \bar{p})$ be the corresponding privacy compromise amount by DH $i$. The set of DHs that embrace positive compromise amounts, i.e., $\{i:\bar{q}_{i} > 0\}$, at the PCE is given by the set $\bar{N} = \{1,2,......,\bar{n}\}$, with an $\bar{n}$ that satisfies 
\begin{equation}
\sum_{i}^{\bar{n}}(C'_{i})^{-1}(C_{\bar{n}}^{0}) \le d \le \sum_{i}^{\bar{n}}(C'_{i})^{-1}(C_{\bar{n} + 1}^{0}). 
\end{equation}
Moreover, benefit $\bar{p}$ at the PCE satisfies
\begin{equation}
C_{\bar{n}^{0}} \le \bar{p} \le C_{\bar{n} + 1}^{0},
\end{equation} 
for any $i \in \bar{N}$, $\bar{p} = C'_{i}(\bar{q}_{i})$. }
\label{theorem2}
\end{theorem}

\noindent \textbf{Theorem Implications} - The theorem states that the PCE has a waterfilling structure - the base privacy compromise cost $C'_{i}(0)$ determines whether DH $i$ compromises privacy or not. The higher the marginal cost at zero, the less likely the DHs will join the privacy compromise program, i.e., embrace a positive amount of compromise. Moreover, the DHs who join the privacy program at PCE bear the same marginal cost. The theorem also implies individual rationality is guaranteed at PCE, i.e., each DH in the privacy compromise program makes non-negative net revenue - we state this as the following corollary, the proof of which is in the Section \ref{proof}. 

\begin{corollary}
\emph{Any DH who participated in the privacy compromise program receives non-negative net revenue at PCE, i.e., $\bar{p}\bar{q}_{i} - C'_{i}(\bar{q}_{i}) \ge 0$ for all $i\in \bar{N}$. }
\label{corollary1}
\end{corollary}

\noindent \textbf{Market `Win-Win' for Ecosystem Stakeholders} - An efficient privacy trading market implies that (a) DHs are led to optimal tradeoffs on how much to compromise aggregate client privacy versus the per-unit compromise (monetary) benefit they get from the ad-network, (b) the ad-network satisfies the downstream demand from the advertisers on their informational requirement, (c) advertisers, through the ad-network can get get their ads placed to the right audience, and (d) consumers, via the monetary benefits received by DHs from the ad-network, either get to pay less for their services, or view fewer ads to improve the QoE. They also see useful targeted ads. 

\subsection{Oligopolistic Markets}
In oligopolistic competition markets, DHs are `benefit anticipating', i.e., the DHs know that the benefit $p$ is set according to (3) and behave strategically. Such markets arise when there are a few DHs in the market \emph{strategically} competing with one another on specific types of consumer information that might be sensitive to the latter (e.g., location, device ID, genetic information). 
We denote the supply function for all DHs but $i$ as $b_{-i} = (b_{1}, b_{2},....,b_{i-1}, b_{i + 1}, .....,b_{|N|})$ and write $(b_{i}, b_{-i})$ for the supply function profile $b$. Each DH $i$ chooses $b_{i}$ to maximize its own benefit $u_{i}(b_{i}, b_{-i})$ given others' bidding strategy $b_{-i}$
\begin{equation}
u_{i}(b_{i}, b_{-i}) = p(b)q_{i}(p(b), b_{i}) - C_{i}(q_{i}(p(b), b_{i})),
\end{equation}
that simplifies to
\[u_{i}(b_{i}, b_{-i}) = \frac{d^{2}b_{i}}{(\sum_{j}b_{j})^{2}} - C_{i}\left(\frac{db_{i}}{(\sum_{j}b_{j})}\right). \]

\noindent Here, the second equality is obtained by substituting the market clearing benefit $p(b) = \frac{d}{\sum_{i}b_{i}}$ and the linear supply bidding function $q_{i}(p(b), b_{i}) = b_{i}p(b)$ into the first equality. As a result functions $\{u_{i}(b_{i}, b_{-i})_{i\in N}$ define a privacy compromise game.

\begin{definition}
\emph{A supply function profile $b^{*}$ is an oligopolistic Nash equilibrium (ONE) if for all DHs $i\in N$, we have
\[u_{i}(b_{i}^{*}, b_{-i}^{*}) \ge u_{i}(b_{i}, b_{-i}^{*}), \, \forall b_{i} \ge 0.\]}
\label{definition2}
\end{definition}

\noindent In order to derive results regarding the existence and uniqueness characteristics of Nash equilibria in oligopoly markets, we first propose the following \emph{three} lemmas (for investigating the existence and uniqueness of ONE), which are proved in the Section \ref{proof}. 

\begin{lemma}
\emph{If $b^{*}$ is an ONE of the privacy compromise game, then $\sum_{j\ne i}b_{j}^{*} > 0$ for any $i\in N$. }
\label{lemma1}
\end{lemma}

\noindent Lemma \ref{lemma1} also directly implies the following lemma, which we state without proof. 

\begin{lemma}
\emph{If $b^{*}$ is an ONE of the privacy compromise game, then at least two DHs have $b_{i}^{*} > 0$. 
\label{lemma2}}
\end{lemma}

\begin{lemma}
\emph{If $b^{*}$ is a Nash equilibrium of the privacy compromise game, then $b_{i}^{*} < B_{-i}^{*} = \sum_{j\ne i}b_{j}^{*}$ for any $i\in N$, and each DH will compromise an amount less than $\frac{d}{2}$ at the ONE, and no ONE exists when $|N|$ = 2.
\label{lemma3}}
\end{lemma}

\noindent The proof of Lemma \ref{lemma3} is in Section \ref{proof}. We now turn to state the \emph{first} of the two main results in this section. 

\begin{theorem}
\emph{Assume that $|N| \ge$ 3. The privacy compromise game has a unique ONE. The ONE solves the following convex optimization problem:\\
$\min_{0\le q_{i} < \frac{d}{2}} \sum_{i}D_{i}(q_{i})$ subject to $\sum_{i}q_{i} = d$, \\
where $D_{i}(q_{i}) = \left(1 + \frac{q_{i}}{d - 2q_{i}}\right)C_{i}(q_{i}) - \int_{0}^{q_{i}}\frac{d}{(d - 2x_{i})^{2}}C_{i}(x_{i})dx_{i}.$}
\label{theorem3}
\end{theorem}

\noindent \textbf{Theorem Implications} - The theorem implies that there exists a pure and unique ONE strategy vector of DH privacy compromise amounts for all DHs at a particular homogeneous ONE benefit $p^{*}$ set by the ad-network that meets the aggregate ad-network demand of $d$ units of total privacy compromise, \emph{but does not provide a guarantee on maximizing utilitarian social welfare amongst the DHs} (see Section \ref{Efficiency_Analysis} in the paper for a mathematical explanation). \emph{In a nutshell, the theorem states that at an oligopolistic privacy trading market between heterogeneous DHs and an ad-network leads to an equilibrium state that is not economically efficient.}
From the proof of the theorem in the Section \ref{proof}, it can be seen as reverse-engineering from ONE to a global optimization problem. Define $\Delta C_{i}(q_{i}) = \frac{q_{i}}{d} - 2q_{i}C_{i}(q_{i}) -  \int_{0}^{q_{i}}\frac{d}{(d - 2x_{i})^{2}}C_{i}(x_{i})dx_{i}$. Then $D_{i}(q_{i}) = C_{i}(q_{i}) + \Delta C_{i}(q_{i})$. Thus, $\Delta C_{i}(q_{i})$ can be interpreted as ``false information" reported by the DHs to gain more benefit from privacy compromise by the ad-network, through strategic bidding. Note that $\Delta_{i}C_{i}(q_{i}) > 0$ for all $q_{i}\in [0, \frac{d}{2})$. $\Delta_{i}C_{i}(q_{i})$ being greater than zero implies that all DHs fake a higher cost function in order to increase the benefit. 

\noindent \textbf{Not the Best `Win-Win' for Ecosystem Stakeholders} - A `no-guarantee' on the efficiency of privacy trading oligopoly implies that DHs might not be able to strategize in a manner so as to converge upon optimal compromise-benefit tradeoffs, but the existence of a unique market equilibrium suggests stable strategizing by the former, i.e., a win-win state that is not the best one. This means that the DHs will fake high costs of compromise to get more benefits that will transfer more incentives to the consumer side at ONE, when compared to PCE. However on the flip side, the privacy compromise amounts at ONE will be higher (not something the DHs would prefer) based on the true compromise costs of the DHs. From a privacy perspective, this result is fairly intuitive as various price strategic mobile apps \emph{sell data that are correlated among the apps, and this correlation negatively affects privacy preservation guarantees at the ad-exchange.} The ad-network and the advertisers are able to satisfy their objectives, as in the PCE. 

\noindent Based on Theorem \ref{theorem3}, similar to the case of perfectly competitive markets, we can further study how a cost function affects a DH's privacy compromise amount at ONE. For each DH $i$, we define the base privacy compromise marginal cost as $C_{i}^{0} = C'_{i}(0^{+})$. Without loss of generality, we assume that  $C_{1}^{0} \le  C_{2}^{0} \le .......\le  C_{|N|}^{0}$. Also notice that $C'_{i}(0^{+}) = D'_{i}(0^{+})$. For modeling convenience, we also introduce parameter $C_{|N| + 1}^{0}$ and set its value to $\max_{i} D'_{|N|}(\frac{d}{3})$. Thus, we have  $C_{1}^{0} \le  C_{2}^{0} \le .......\le  C_{|N|}^{0} \le C_{|N| + 1}^{0}$. We now have the \emph{second} important result (see Section \ref{proof} for a proof) for this section, on privacy compromise characteristics of DHs. 

\begin{theorem}
\emph{Let $|N| > 3$, $\{(b_{i}^{*})_{i\in N}\}$ be an ONE, $p^{*} = \frac{d}{\sum_{i}b_{i}^{*}}$ be the ONE benefit, and $q_{i}^{*} = b_{i}^{*}p^{*}$ be the corresponding privacy compromise amount by DH $i$. The set of DHs $i$ that embrace positive compromise amounts, i.e., $\{i: q_{i}^{*} > 0\}$, at the ONE is given by the set $N^{*} = \{{1,2,......,n^{*}}\}$, with an $n^{*}$ that satisfies
\begin{equation}
\sum_{i}^{n^{*}}(D'_{i})^{-1}(C_{n^{*}}^{0}) \le d \le \sum_{i}^{n^{*}}(D^{'}_{i})^{-1}(C_{n^{*} + 1}^{0})
\end{equation}
Moreover, benefit $p^{*}$ at the ONE satisfies
\begin{equation}
C_{n^{*}}^{0} \le p^{*} \le C_{n^{*} + 1}^{0},
\end{equation} 
for any $i \in N^{*}$, $p^{*} = D'_{i}(q_{i}^{*})$. }
\label{theorem4}
\end{theorem}

\noindent \textbf{Theorem Implications} - The theorem states that the ONE has a waterfilling structure, and henceforth the implications are exactly the same as for Theorem \ref{theorem2}. The theorem also implies individual rationality is guaranteed at ONE, i.e., each DH in the privacy compromise program makes non-negative net revenue - we state this as the following corollary, the proof of which is in the Section \ref{proof}. 

\begin{corollary}
\emph{Any DH who participated in the privacy compromise program receives non-negative net revenue at ONE, i.e., $p^{*}q_{i}^{*} - C'_{i}(q_{i}^{*}) \ge 0$ for all $i\in N^{*}$.} \label{corollary2}
\end{corollary}

\section{Efficiency and Optimality Aspects} \label{Efficiency_Analysis}
In this section, we characterize efficiency loss of oligopoly privacy trading markets and derive the optimality of our mechanism choice. 
\subsection{Characterizing Efficiency Loss at ONE}
We have shown that utilitarian social welfare is maximized at PCE, thereby making perfectly competitive markets efficient. In contrast, due to DHs' benefit-anticipating and strategic behavior, the ONE is expected to be less efficient. In this section, we investigate the efficiency loss at ONE for different degrees of heterogeneity among DH cost functions, and provide closed form characterization of the efficiency loss (if any). Here, we define the the efficiency loss as the ratio of the total disutility at PCE to the minimum total disutility, i.e., the ratio $\frac{C^{*}}{C}$. \emph{Thus, efficiency loss is equivalently the price of anarchy (PoA) \cite{roughgarden2005selfish}.} To this end, we have the following main result post investigation.

\begin{theorem}
\emph{Let $\{(\bar{b}_{i})_{i\in N}, \bar{p}\}$ be a perfectly competitive equilibrium (PCE), and $p^{*}$ be the corresponding benefit at the oligopolistic Nash equilibrium (ONE). We have the following: 
\begin{enumerate}
\item $\bar{N} \subseteq N^{*}$ where $\bar{N}$ is the set of DHs who participate in the privacy compromise program at PCE, and $N^{*}$ is the set of DHs who participate in the privacy compromise program at ONE. 
\item $\bar{p} \le p^{*} \le n - \frac{1}{n} - \frac{2M}{m\bar{p}}$, where $M = \max_{i\in N}C'_{i}(\frac{d}{n})$; $m = \min_{i\in N}C'_{i}(\frac{d}{n})$.
\item $\bar{C} \le C^{*}$, and if we assume that $\bar{q}_{\max} = \max_{i}\bar{q_{i}} < \frac{d}{2}$, then we have 
\[C^{*} \le (1 + \frac{\bar{q}_{\max}}{d} - 2\bar{q}_{\max})\bar{C},\]
where $\bar{C} = \sum_{i}C_{i}(\bar{q}_{i})$ be the total social cost at PCE, and  $C^{*} = \sum_{i}C_{i}(q_{i}^{*})$ is the total social cost at ONE. 
\end{enumerate}}
\label{theorem5}
\end{theorem}

\noindent \textbf{Theorem Implications} - The conditions in the theorem together imply the following:
\begin{itemize}
\item The set of DHs that contribute to the privacy compromise program at ONE is a superset (due to more DHs seeing an opportunity to make benefits by bidding strategically) of that at PCE (due to the non-strategic nature of the DHs at PCE). 
\item The benefit at the ONE is higher than that at PCE (due to strategic DH behavior at ONE), but the ratio between the two benefits are bounded. This last point makes sure that there are limits of DHs to exploiting the advantage of strategic behavior over non-strategic behavior. 
\item The total (aggregate) compromise cost at the ONE is higher than that at the PCE (due to strategic higher bidding, consequently more benefits, consequently unwanted additional privacy compromise), but the ratio between the two costs are bounded (incentivizing strategic higher bidding over non strategic bidding), provided no one compromises more than half of the total demand at the PCE (can be enforced via regulation). 
\item In addition, as long as no DH compromises more than $\frac{d}{3}$ at PCE, the efficiency loss $\frac{C^{*}}{C}$ is bounded by $\frac{3}{2}$. This condition can be guaranteed if there are at least three DHs having comparably low compromise cost (e.g., big firms with a huge base of locked-in clients and/or firms trading non-sensitive data), compared to the others. The presence of closed form expressions for the efficiency loss may serve as a guideline to regulators for limiting the market power of some DHs (in the oligopoly setting) to maximize social welfare (e.g., by allowing the entry of new moderate/big DH app firms in the market to stiffen competition, and/or control types of data to be traded).
\end{itemize}
Moreover, from Theorems \ref{theorem2} and \ref{theorem4}, we can derive the following special case result if the DHs have homogeneous costs, and the difference between the two market equilibria, i.e., PCE and ONE, are small. The proof of the result is in the Section \ref{proof}. 
\begin{corollary}
\emph{On the condition that DHs have the same cost function, we have the following:
%\begin{enumerate}
1. $p^{*} = n - \frac{1}{n} - 2\bar{p}$. As $n \rightarrow \infty$, $p^{*} \rightarrow \bar{p}$.  2. $C^{*} = \bar{C}$. As $n\rightarrow \infty$, $C^{*} \rightarrow \bar{C}$.} 
%\end{enumerate}
\label{corollary3}
\end{corollary}

\noindent The condition guarantees that when app firms facing similar cost structure (due to trading similar data type) are in competition, applying the supply function bidding scheme will lead to system efficiency irrespective of whether the market is perfectly competitive or oligopolistic.   

\noindent \textbf{Can the Efficiency Loss be Unbounded?} - We show with an example that the efficiency loss in the worst case can be unbounded. Consider the case where there are three DHs with cost functions $C_{1}(q) = \frac{1}{2rcq^{2}}$, and $C_{2}(q) = C_{3}(q) = \frac{1}{2cq^{2}}$, where $c$ and $r$ are constant parameters. Using Theorem \ref{theorem2}, we can calculate the PCE to be: $\bar{q}_{1} = \frac{r}{r + 2d}$, $\bar{q}_{2} = \bar{q}_{3} = \frac{1}{r + 2d}$, and $\bar{p} = \frac{r}{r + 2cd}$. Similarly, using Theorem \ref{theorem3}, we get the ONE as: $q_{1}^{*} = \frac{-r + \sqrt{(16 + 9r)r}}{4(2 + r)d}, q_{2}^{*} = q_{3}^{*} = \frac{8 + 5r - \sqrt{(16 + 9r)r}}{8(2 + r)d}$, and $p^{*} = D - \frac{q_{1}^{*}}{D} - 2q_{1}^{*}q_{1}^{*}$. Now let $r \rightarrow \infty$ - for the PCE we then have $\bar{q}_{1} \rightarrow d$, $\bar{q_{2}}, \bar{q_{3}} \rightarrow 0$, $\bar{p} \rightarrow cd$, and total cost $\bar{C} \rightarrow 0$. For the ONE, we have $q_{1}^{*} \rightarrow \frac{d}{2}$, $q_{2}^{*}, q_{3}^{*} \rightarrow \frac{d}{4}$, $p^{*} \rightarrow \infty$, and the total cost $C^{*} \rightarrow \frac{cd^{2}}{4}$. Thus, $\frac{p^{*}}{p} \rightarrow \infty$, and $\frac{C^{*}}{C} \rightarrow \infty$. 

\noindent \textbf{Message for Regulators} - We see that in a market with DHs having extremely heterogeneous cost functions, the efficiency loss at the ONE might be unbounded. Combining this fact with the implications of Corollary \ref{corollary3}, regulators are advised to enable privacy trading by apps in segregated pools, with similar data types to be traded.

\subsection{Optimality of Our Mechanism Choice}\label{s4.4}
 We prove the optimality of our mechanism choice, i.e., a linear supply function mechanism, over a class of mechanisms that are suited to designing markets for our problem. 

\noindent To embark on this task, we first consider a mechanism desirable if it minimizes worst case efficiency loss when DHs are `benefit anticipating', independent of the utility functions of the DHs and their number. That is, the mechanisms we seek are those that perform well under broad assumptions of the nature of the preferences of the market participants. We will show that under a specific set of assumptions, our mechanism choice minimizes the worst case efficiency loss when compared to all other feasible mechanisms fitting the assumptions. To this end, we first define the class, $\mathcal{M}$, of mechanisms that we want to consider.

\begin{definition}
\emph{The class $\mathcal{M}$ of mechanisms consists of all supply functions, $M(b, p)$, such that the following conditions are satisfied:
\begin{enumerate}
    \item $M$ defines a smooth market-clearing mechanism. Here, a differentiable $M:(0, \infty) \times \mathrm{R}^{+} \rightarrow \mathrm{R}^{+}$ is said to be a smooth market clearing mechanism if for all $d > 0$, for all $n=|N| > 1$, and for all non-zero $b = (b_{1},..., b_{N})$, $\exists$ a unique solution $p > 0$ to
    \begin{equation}
        \sum_{i}^{n}M(b_{i}, p) = d. 
    \end{equation}
    \item For all $C_{i} \in \mathcal{C}$, for all $u \in \mathcal{U}$, and for all $d > 0$, a DH's payoff is concave if it is benefit anticipating. $\mathcal{C}$ is the set consisting of all continuous, convex, and strictly increasing cost functions.
    \item For all $C_{i} \in \mathcal{C}$, for all $u \in \mathcal{U}$, and for all $d > 0$, there exists a $b \ge 0$ such that $M(b_{i}, p)$ = $q_{i}(b_{i}, p)$, $\forall i.$ 
\end{enumerate}}
%\ref{definition3}
\end{definition}
The second condition allows us to characterize Nash equilibria in terms of only the first-order conditions. To justify this condition, we note that some assumption of quasiconcavity is generally used to guarantee the existence of pure-strategy Nash equilibria \cite{mas1995microeconomic}. The third condition ensures that given a benefit $p$ and given $q_{i}(b_{i}, p) \in [0, d]$, each DH $i$ can make a choice $b_{i}$ to guarantee $q_{i}(b_{i}, p)$ - ensuring all possible demands can be chosen any market-clearing benefit. \emph{In view of these conditions, it is evident that the class of mechanisms in $\mathcal{M}$ fit the privacy trading scenario we address in this work.} In this regard, we showcase the optimality of our proposed parametric mechanism, an element of the set $\mathcal{M}$, via the following theorem, the proof of which is in the Section \ref{proof}.
\begin{theorem}
\emph{Given $M \in \mathcal{M}$, the following results hold:
\begin{enumerate}
    \item There exists a competitive equilibrium $b$ for any privacy trading market characterized by the triplet $(d, N, U)$, where $d$ is the total privacy compromise demand on the ad-network side, $N$ is the number of competing DHs, and $U$ is the vector of utility functions for every DH. Moreover, for any such $b$, the resulting privacy compromises, $q_{i}(b_{i}, p)$, for each DH $i$ maximizes welfare.
    \item There exists $B:(0, \infty) \rightarrow (0, \infty)$, a concave, strictly increasing, differentiable, and invertible function, such that for all $p > 0$, and $b_{i} \ge 0, \forall i\in N$, we have $M(b_{i},p) = b_{i}B(p)$. 
    \item The worst case market efficiency loss under oligopoly is minimized if $M(b_{i},p) = \Delta b_{i}p$, for some $\Delta > 0$. 
\end{enumerate}}
\label{theorem10}
\end{theorem}
\noindent \textbf{Theorem Implication} - For privacy trading oligopoly markets, the linear supply function mechanism minimizes the loss in worst case market efficiency.

% -*- ispell-local-dictionary: "american"; TeX-master: "cloudshare.tex"; -*-
%
\section{Computational Evaluation}\label{Algo_Bidding}
In this section, we focus on developing supply function bidding algorithms that converge in practice to market equilibria for perfectly competitive and oligopolistic markets in a distributed fashion. Our primary performance metric is market equilibrium convergence speed in terms of the number of iterations. Our motivation for coming up with distributed algorithms is the fact that DH cost functions are private information not released to an ad-network, and as a result the latter cannot centrally solve the optimization problems to maximize utilitarian social welfare and arrive at ONE, respectively. In addition, we need algorithms that are light on computation and communication overhead. 

% \vspace{-1 mm}
\subsection{Mini Real-World Evaluation Setup}
 As part of a mini-experiment to evaluate supply function bidding algorithms, we collect sanitized consumer data for 1000 clients on their two sleep patterns (i.e., time to go to sleep, hours of sleep) from three fitness app startup firms A, B, and C based in northern California, USA. We ensure that the set of 1000 clients for each company do not overlap. For the aggregate data collected from both the companies, we set up an independent (of A, B, and C) sleep expert representative from a medical department at an university in northern California to \emph{act} as an ad-network. The expert has thirty years of experience in research and consulting, and more importantly possesses deep knowledge of what type of sleep data would be of interest to different commercial organizations in the fitness and pharmaceutical industries. Having collected real-world data, as a mock experiment, we synthetically implement a triopoly competition between A, B, and C by choosing a senior representative from both the firms to trade on the sanitized data of their clients with the ad-network, i.e., the medical representative, in return for (a) fictitious (but scaled on medical value of the data) monetary benefits and (b) some health insights on the available consumer data to be passed on by the representatives of A, B, and C to their clients. We emphasize here that the ad-network does not have knowledge of individual consumers whose data is under trade. Trading is done using the supply function mechanism and each of A, B, and C choose parameters of 1, 1, and 2 respectively, with a common demand upper limit of 100 differential privacy (DP) units, and a zero lower limit. Each DP unit is assumed to be 0.02. Each DH reports a nearly linear cost function to be of the form $C_{i}(q_{i}) = a_{i}q_{i} + h_{i}q_{i}^{2}$ with $a_{i} \ge 0$ and $h_{i} << a_{i}\ge 0$. More specifically, $a_{i}$ values chosen by firms A, B, and C are 0.1, 0.2, and 0.1 respectively. Correspondingly, the $h_{i}$ values chosen are 0.002, 0.005, and 0.005 respectively.
% \vspace{-3 mm}
\begin{figure*}[h!]
  \centering
\begin{minipage}{0.45\textwidth}
  \centering
    \includegraphics[angle=-90,width=1\linewidth]{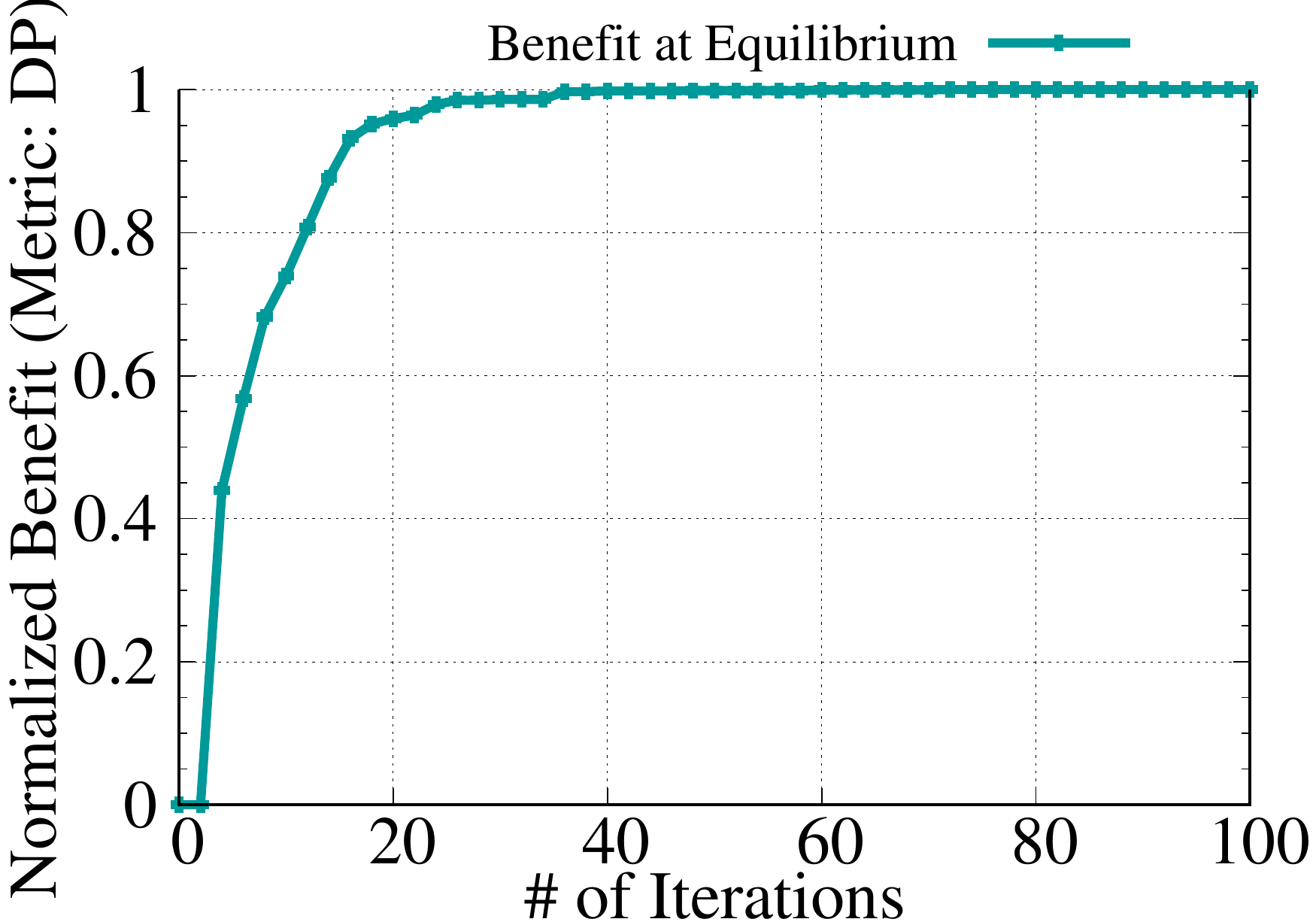}
\end{minipage}\;
\begin{minipage}{0.45\textwidth}
  \centering
    \includegraphics[angle=-90,width=1\linewidth]{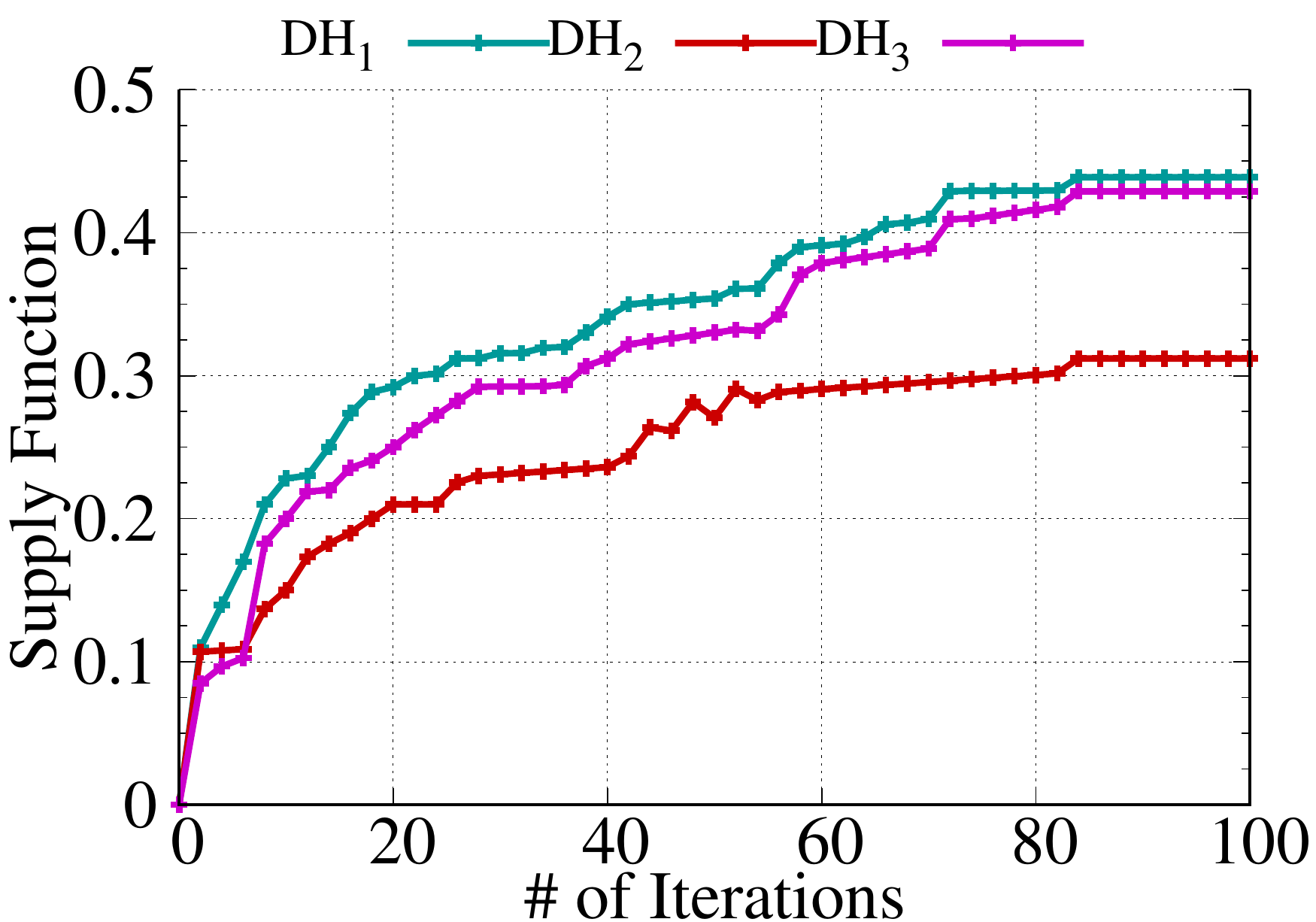} 
\end{minipage}
%\end{subfigure}
\caption{\small (Benefit, Supply Function) at Market Equilibrium}\label{fig:00a}
\end{figure*}
% \vspace{-5 mm}

\begin{figure}[h!]
\begin{center}
    \includegraphics[width=0.9\linewidth]{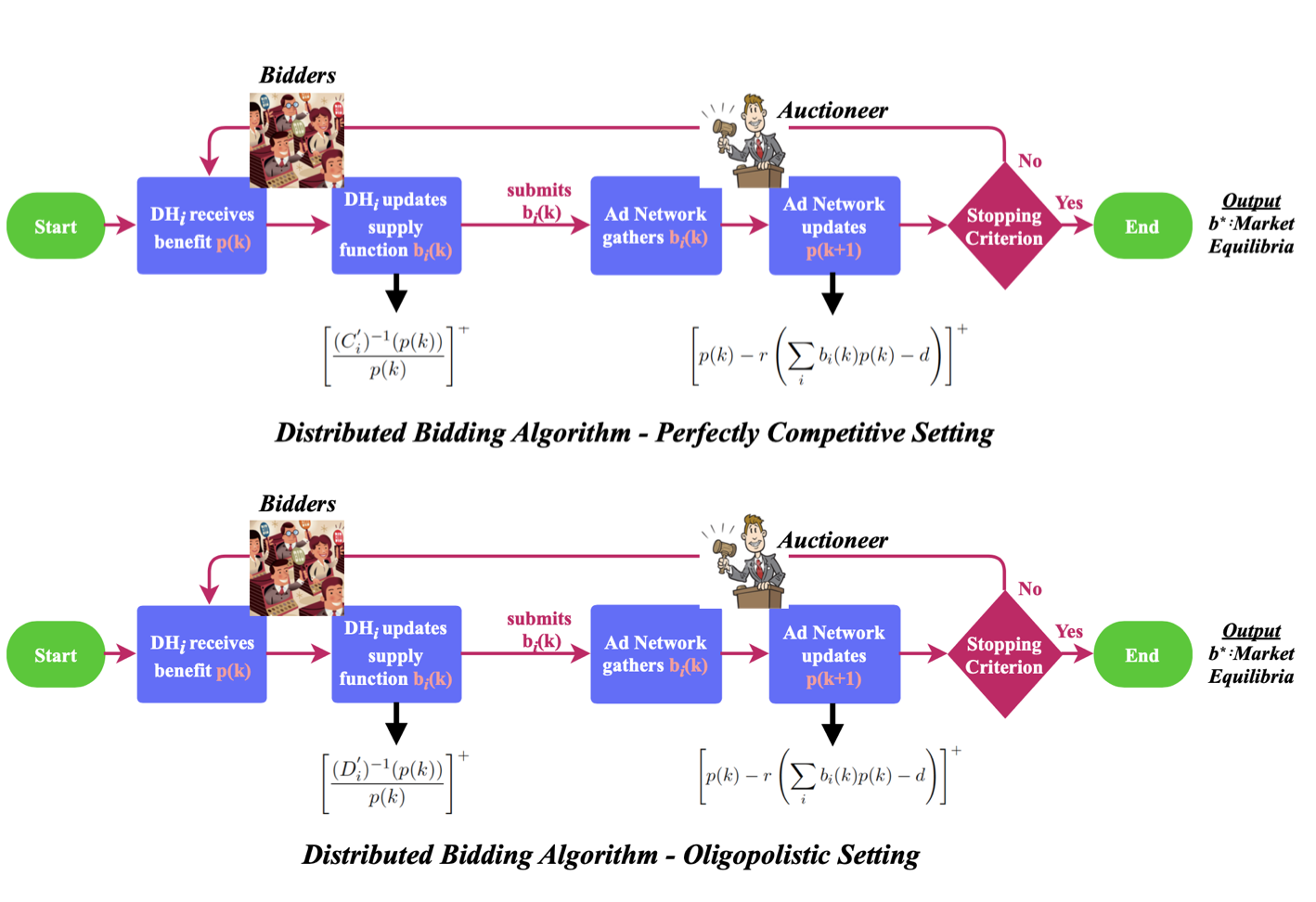}\par
    % \vspace{-3 mm}
\caption{Flowchart of Distributed Market Bidding Algorithms}\label{...}
\end{center}
% \vspace{-5 mm}
\end{figure}
% \vspace{-3 mm}

\subsection{Distributed Bidding Algorithms}
As potential distributed algorithm candidate types, one could either use the standard dual gradient algorithm proposed in \cite{bertsekas1989parallel}, or the alternative direction multiplier method in \cite{boyd2011distributed}. Both types are iterative in nature, and equivalently maps the supply bidding process. In this work\footnote{We do not focus on the design of optimal distributed algorithms in terms of speed and scalability. Our goal is to just show fast convergence and scalability promise of implemented markets induced by supply function theory, and our proposed algorithms achieve them using as basis, the seminal algorithm type in \cite{bertsekas1989parallel}\footnote{This type of an algorithm is an example of standard and widely popular \emph{t\^atonnement process} \cite{varian1992microeconomic}\cite{arrow1981handbook} to converge to market equilibrium in a computational manner.}.} we resort to the dual gradient algorithm in \cite{bertsekas1989parallel}, without loss of generality. \emph{The basic idea behind the two algorithms (see Algorithms \ref{algorithm1} and \ref{algorithm2} for perfectly competitive and oligopolistic markets, respectively) is the iterative interplay (until convergence) between the ad-network announcing a benefit $p$ to the DHs, and the DHs subsequently updating their non-private bidding functions $b_{i}$ to the ad-network.} (see Figure 4 for a flowchart representation) In principle, the crux lies behind convergence lies in the Lagrangian of Equation (7) being strictly concave and thereby using the Projection Theorem \cite{boyd2004convex} we arrive at the optimal benefit and supply functions at market equilibrium. 
Consequently, our proposed distributed bidding algorithms possess all the convergence properties of dual gradient algorithms. We refer the readers to \cite{bertsekas1989parallel} for details regarding the theory of optimal step sizes, the stopping criterion, and convergence speed. As an example of the high convergence speed, we show via experiments in the following section that for very low $\gamma$ values in Algorithms \ref{algorithm1} and \ref{algorithm2}, convergence is very fast, thereby showing great potential to ensure the property of scalability for large number of DHs. To be more specific, it is shown in \cite{bertsekas1989parallel} that in theory very small $\gamma$ values result in an exponential convergence rate. 

% \vspace{-2 mm}

\subsection{Evaluation Results}
For our real-world experimental setting, we show in Figure 3 the results for benefit and supply function values at market equilibrium with respect to the number of iterations to market convergence. We observe that benefit and supply functions converge fast (within 25 iterations on a latest MacBook Pro with 16GB RAM) to the market equilibrium (ONE). This indicates the possibility of the existence of working markets satisfying all concerned stakeholders (as per our model) if personal data were to be traded. \emph{As part of future plans, we would like to run larger scale field experiments, conditioned on the availability of real data, to validate our speed and scalability claims on working privacy trading markets.} However, in the absence of real-world data, we experiment with synthetic data as curated in Section VI.C. \emph{Without loss of generality (and in the interest of space), we represent one of the 50 random instances in our plots. We not later on the rationale of not showing confidence interval bars in the plots.}  
 
% \vspace{-2 pt}
%\hline
\begin{algorithm}
        \caption{Distributed Bidding Algorithm - Perfectly Competitive Setting}
        \label{algorithm1}
        %\hline
        \begin{algorithmic}[1]
            \STATE On receiving benefit $p(k)$ announced by the ad-network, 
            each $DH_{i}$ updates its supply function, 
            $b_{i}(k)$ according to
	\begin{equation}\label{alg1_competitive}
		b_{i}(k) = {\left[\frac{(C^{'}_{i})^{-1}(p(k))}{p(k)}\right]}^{+}
	\end{equation}and submits it to the ad-network. Here ``+'' denotes the projection onto ${\mathbf{R}^{+}}$, the set of non-negative real numbers.
            \STATE On gathering bids $b_{i}(k)$ from DHs, the ad-network updates the benefit according to
	\begin{equation}\label{alg2_competitive}
		p(k+1)  = {\left[p(k) - r\left(\sum_{i}b_{i}(k)p(k) - d\right)\right]}^{+}
	\end{equation}and announces the benefit $p(k+1)$ to the DHs, where $r > 0$ is a constant stepsize.
     \STATE Set $k \rightarrow k+1$
      \STATE Check stopping criterion as mentioned in \cite{bertsekas1989parallel}, and repeat
        \end{algorithmic}
    \end{algorithm}
% \vspace{-10 pt}
\begin{algorithm}
        \caption{Distributed Bidding Algorithm - Oligopolistic Setting}
        \label{algorithm2}
        \begin{algorithmic}[1]
            \STATE On receiving benefit $p(k)$ announced by the ad-network, each $DH_{i}$ updates its supply function, $b_{i}(k)$ according to
	\begin{equation}\label{alg1_oligopolistic}
		b_{i}(k) = {\left[\frac{(D^{'}_{i})^{-1}(p(k))}{p(k)}\right]}^{+}
	\end{equation}and submits it to the ad-network. Here ``+'' denotes the projection onto ${\mathbf{R}}^{+}$, the set of non-negative real numbers.
            \STATE On gathering bids $b_{i}(k)$ from DHs, the ad-network updates the benefit according to
	\begin{equation}\label{alg2_oligopolistic}
		p(k+1)  = {\left[p(k) - r\left(\sum_{i}b_{i}(k)p(k) - d\right)\right]}^{+}
	\end{equation}and announces the benefit $p(k+1)$ to the DHs, where $r > 0$ is a constant stepsize.
     \STATE Set  $k \rightarrow k+1$
      \STATE Check stopping criterion as mentioned in \cite{bertsekas1989parallel}, and repeat
        \end{algorithmic}
    \end{algorithm}

\section{Summary and Future Work}
In this paper, we proposed a introductory but rigorous preference-based privacy trading market model for mobile in-app ecosystems of the current data surveillance age that aims to achieve a maximum privacy welfare state amongst competing data holders (e.g., apps) by preserving their \emph{heterogeneous} privacy preservation constraints upto certain compromise levels (in return for benefits to data holders), induced by their clients, and at the same time satisfying requirements of agencies (e.g., advertisers) that collect client data for the purpose of targeted advertising. \emph{More importantly, our proposed trading methodology is consensual in the sense that pre-trading, DHs can decide on their trading preferences as a function of the benefit to be offered, without needing to sell non-voluntarily with no explicitly offered benefit.} 
To this end, using concepts from supply-function economics, we proposed the first mathematically rigorous privacy market design paradigm with private DH cost functions that characterized states of market efficiency as well as inefficiency by respecting \emph{heterogeneous privacy constraints} of competing data holders to extents possible, in a provably optimal fashion. More specifically, we analyzed perfectly competitive and oligopolistic markets to achieve market equilibria that is efficient in the former, but not in the latter due to negative externalities of trading not being internalized. Consequently, we characterized the efficiency gap in closed form. As a major finding, we showed that increasing competition between app firms of similar market power for privacy trading activities contribute to increased economic social welfare due to trading externalities being internalized better between similar firm types, thereby suggesting regulators to enable privacy trading in segregated pools of similar app firms. 

\noindent As part of future work, we plan to (a) gauge the preference supply functions of individual DHs using large-scale social experiments, and (b) investigate the existence of efficient/boundedly inefficient multi-supplier (apps), multi-demand side (ad-exchanges) market competition models in a privacy trade setting, and explicitly account for information correlations between supplier side data. 

\section{Proofs of Theorems}
\label{proof}
Following the development in \cite{li2015demand} - developed atop \cite{johari2011parameterized} (in the main document), we state the following proofs to the results in our work using similar notation (due to appropriate variable-meaning characterization).

\noindent \textbf{Proof of Theorem 1:} Definition 1 tells that $\{(\bar{b}_{i})_{i \in N}, \bar{p}\}$ is a competitive equilibrium if and only if

%\begin{equation}\label{eq1}
%(C^{'}_{i}(q_{i}(\bar{b}_{i}, \bar{p}_{i})) - \bar{p})(b_{i} - \bar{b}_{i}) \geq 0, \; \forall b_{i} \geq 0 
%\end{equation}
%\begin{equation}\label{eq2}
%\sum_{i} {q}_{i}(\bar{b}_{i}, \bar{p}) = d
%\end{equation}

\begin{subequations}
% \label{eq14}
\begin{align}
(C^{'}_{i}(q_{i}(\bar{b}_{i}, \bar{p}_{i})) - \bar{p})(b_{i} - \bar{b}_{i}) \geq 0, \; \forall b_{i} \geq 0  \label{eq14a}\\
\sum_{i} {q}_{i}(\bar{b}_{i}, \bar{p}) = d\label{eq14b}
\end{align}
\end{subequations}

Here, (\ref{eq14a}) results from the optimality condition of the convex optimization problem of DH net revenue, and (\ref{eq14b}) follows directly from Definition 1.
Since $\bar{p} \geq 0$,  multiplying $\bar{p}$ to (\ref{eq14a}), we get

%\begin{equation}\label{eq3}
%(C^{'}_{i}\left(\bar{q}_{i})- \bar{p}\right)\left(q_{i} - \bar{q}_{i}\right) \geq 0, \; \forall q_{i} \geq 0
%\end{equation}

%\begin{equation}\label{eq4}
%\sum_{i} \bar{q}_{i} = d
%\end{equation}

\begin{subequations}
\label{eq15}
\begin{align}
(C^{'}_{i}\left(\bar{q}_{i})- \bar{p}\right)\left(q_{i} - \bar{q}_{i}\right) \geq 0, \; \forall q_{i} \geq 0  \label{eq15a}\\
\sum_{i} \bar{q}_{i} = d \label{eq15b}
\end{align}
\end{subequations}

This is just the KKT optimality condition of the optimization problem in the theorem. Hence, $({q}_{i})_{i \in N}$ maximizes social welfare. And if $\{(\bar{q}_{i})_{i \in N}, \bar{p}\}$ is an optimal solution of the latter optimization problem, $\{\left(\bar{b}_{i} =  \frac{\bar{q}_{i}}{\bar{p}}\right)_{i \in N}, \bar{p}\}$ 
satisfies (\ref{eq14a}) ; 
this tells that  $\{(\bar{b}_{i})_{i \in N}, \bar{p}\}$ is a competitive equilibrium. If $C_{i}(q_{i})$ is convex for each DH $i$, then the social welfare maximization problem is a strictly convex problem. Thus there exists a unique optimal solution $(\bar{q}_{i})_{i \in N}$. Moreover, from (\ref{eq15a}), $\bar{p} = C^{'}_{i}(\bar{q}_{i})$ for any $\bar{q}_{i} \geq 0$ $\Rightarrow \bar{p}$ is unique $\Rightarrow$ unique equilibrium. 
$\blacksquare$
%\subsection{Proof of Theorem 2.}

\noindent\textbf{Proof of Theorem 2:} From the proof of Theorem 1, we know that $\{\bar{p}, (\bar{q}_{i})_{i \in N}\}$ satisfies (\ref{eq15a}) and (\ref{eq15b}). From (\ref{eq15a}), we know that, for any $i \in N$,  1) if $\bar{q}_{i}  > 0$, then $\bar{p} = C^{'}_{i}(\bar{q}_{i}) \geq C^{'}_{i}(0)$, 2) if $\bar{q}_{i}  = 0$, then $\bar{p} \leq C^{'}_{i}(\bar{q}_{i}) = C^{'}_{i}(0)$. Thus, we know all the DHs who compromise on privacy have a smaller $C^{\ast}_{i} = C^{'}_{i}(0)$ than those who do not. Since  $C^{\ast}_{i}$ is increasing in $i$, $\bar{N}$ takes the form of $1, 2, ..., \bar{n}$. If $\bar{n} < |N|$, then 1 and 2 imply that $C^{0}_{\bar{n}} \leq \bar{p} \leq C^{0}_{\bar{n}+1}$. If $\bar{n} = |N|$, $\bar{p} = C^{'}_{|N|}(\bar{q}_{|N|}) \leq C^{'}_{|N|}(d) = C^{0}_{n+1}$, thus $C^{0}_{\bar{n}} \leq \bar{p} \leq C^{0}_{n+1}$. Note that, $C^{'}_{i}(q^{'}_{i})$ is an increasing function. Hence $\sum_{i}^{\bar{n}}(C^{'}_{i})^{-1}(C^{0}_{\bar{n}}) \leq \sum_{i}^{\bar{n}}(C^{'}_{i})^{-1}(\bar{p}) \leq \sum_{i}^{\bar{n}}(C^{'}_{i})^{-1}(C^{0}_{\bar{n}+1})$ which is $\sum_{i}^{\bar{n}}(C^{'}_{i})^{-1}(C^{0}_{\bar{n}}) \leq \sum_{i}^{\bar{n}}\bar{q}_{i} = d \leq \sum_{i}^{\bar{n}}(C^{'}_{i})^{-1}(C^{0}_{\bar{n}+1})$. $\blacksquare$

%\subsection{Proof of Corollary 1.}

\noindent\textbf{Proof of Corollary 1:} From Theorem 2, we know that $\forall i \in \bar{N}$, $\bar{p} = C^{'}_{i}(\bar{q}_{i})$. Notice that $C_{i}(\cdot)$ is a convex function. Thus $C_{i}(\bar{q}_{i}) - C_{i}(0) \leq C^{'}_{i}(\bar{q}_{i})\bar{q}_{i}$. As $C_{i}(0) = 0$, we have $C_{i}(\bar{q}_{i}) \leq \bar{p}\bar{q}_{i}$. $\blacksquare$
%\subsection{Proof of Lemma 1.}

\noindent\textbf{Proof of Lemma 1}: We prove the result by contradiction. Suppose that it does not hold, and without loss of generality, assume that $\sum_{j \neq 1}b^{\ast}_{j} = 0$ for $DH_i$. Then the payoff for the $DH_{i}$ is $U_{i}(b^{\ast}_{i}, b^{\ast}_{-i}) = 0$ if $b^{\ast}_{i} = 0$, and $U_{i}(b^{\ast}_{i}, b^{\ast}_{-i}) = \frac{d^2}{b^{\ast}_{i}} - C_{i}(d)$ if $b^{\ast}_{i} > 0$.  We see that when $b^{\ast}_{i} = 0$, $DH_{i}$ has an incentive to increase it, and when $b^{\ast}_{i} \geq 0$, $DH_{i}$ has an incentive to decrease it. So, there is no Nash equilibrium with $\sum_{j \neq i}b^{\ast}_{j} = 0$. $\blacksquare$
\\
%\subsection{Proof of Lemma 3.}
\noindent\textbf{Proof of Lemma 3}: We have 
\begin{equation} 
\begin{array}{ll}
\label{eq5}
U_{i}(b_{i}, b_{-i}) = p(b)q_{i}(p(b), b_{i}) - C_{i}(q_{i}(p(b), b_{i}))  \\ =\frac{d^{2}b_{i}}{\left(\sum_{j}b_{j}\right)^{2}} - C_{i}\left(\frac{d{b_{i}}}{\sum_{j}b_{j}}\right)
\end{array}
\end{equation}
From (\ref{eq5}), we have
\begin{equation} 
\footnotesize{
\label{eq6}
\begin{split}
\frac{\partial U_{i}(b_{i}, b_{-i})}{\partial b_{i}} = \frac{d^2\left(B_{-i} - b_{i}\right)}{\left(B_{-i} + b_{i}\right)^{3}} - \frac{dB_{-i}}{\left(B_{-i} + b_{i}\right)^{2}}C^{'}_{i}\left(\frac{db_{i}}{B_{-i} + b_{i}}\right) \\
= \frac{d^2}{(B_{-i}+b_{i})^2}\left[\frac{B_{-i} - b_{i}}{B_{-i} + b_{i}}-\frac{B_{-i}}{d}C^{'}_{i}\left(\frac{db_{i}}{B_{-i}+b_{i}}\right)\right]
\end{split}
}
\end{equation}

The first form in the square bracket in (\ref{eq6}) is no greater than 1 and strictly decreasing in $b_{i}$, the second term is increasing in $b_{i}$. So, if $\frac{B_{-i}}{dC^{'}_{i}(0)} \geq 1$ and $\frac{\partial U_{i}(b_{i}, b_{-i})}{\partial b_{i}} \leq 0$ $\forall b_{i}$, and $b_{i} = 0$ maximizes ${DH_{i}}^{'}s$ payoff $U_{i}(b_{i}, b_{-i})$ for the given $b_{-i}$. If $\frac{B_{-i}}{dC^{'}_{i}(0)} \leq 1$,$ \frac{\partial U_{i}(b_{i}, b_{-i})}{\partial b_{i}} = 0$ only at one point $b_{i} > 0$. Furthermore, note that $\frac{\partial U_{i}(0, b_{-i})}{\partial b_{i}} >0$ and $\frac{\partial U_{i}(B_{-i}, b_{-i})}{\partial b_{i}} \leq 0$. So, the point $b_{i}$ maximizes ${DH_{i}}^{'}s$ payoff $U_{i}(b_{i}, b_{-i})$ for a given $b_{-i}$. Thus, at Nash equilibrium, $b^{\ast}$, 

\begin{gather}
\begin{aligned} 
\label{eq7}
%\frac{dU_{i}(b_{i}, b_{-i})}{db_{i}} = \frac{d^2\left(B_{-i} - b_{i}\right)}{\left(B_{-i} + b_{i}\right)^{3}} - \frac{dB_{-i}}{\left(B_{-i} + b_{i}\right)^{2}}C^{'}_{i}\left(\frac{db_{i}}{B_{-i} + b_{i}}\right) = \frac{d^2}{(B_{-i}+b_{i})^2}\left[\frac{B_{-i} - b_{i}}{B_{-i} + b_{i}}-\frac{B_{-i}}{d}C^{'}_{i}\frac{db_{i}}{B_{-i}+b_{i}}\right]
%If \frac{B&{\ast}_{-i}}{dC^{`}_{i}(0)} \geq 1
  b^{\ast} satisfies
  \begin{cases}
   b^{\ast}_{i} = 0, \forall i, \text{if $\frac{B^{\ast}_{-i}}{dC^{'}_{i}(0)} \geq 1$}  \\
    \frac{B^{\ast}_{-i} -  b^{\ast}_{i}}{B^{\ast}_{-i} +  b^{\ast}_{i}} - \frac{B^{\ast}_{-i}}{d}C^{'}_{i}\left(\frac{db^{\ast}_{i}}{B^{\ast}_{-i} +  b^{\ast}_{i}}\right) = 0, \text{otherwise} 
  \end{cases}
\end{aligned}
\raisetag{30pt}
\end{gather}
Given a Nash equilibrium, $b^{\ast}$: 1) if $b^{\ast}_{i} =0$, then $b^{\ast}_{i} < B^{\ast}_{-i}$ from lemma 1 and, 2) otherwise, $b^{\ast}_{i}$ satisfies (\ref{eq7}). Note that the second term on the left hand side of (\ref{eq7}) is positive. So the first term must be positive as well, which requires $B^{\ast}_{-i} > b^{\ast}_{i}$. Because for each $DH_{i}$, $q^{\ast}_{i} = \frac{b^{\ast}_{i}d}{b^{\ast}_{i}} + B^{\ast}_{-i}$, each DH will compromise a privacy of less than $\frac{d}{2}$ at the equilibrium.  $\blacksquare$

%\subsection{Proof of Theorem 3.}
\noindent\textbf{Proof of Theorem 3:} Here, we prove the existence and uniqueness of the optimal solution of optimization problem in Theorem 3. We first pick $\hat{d} < \frac{d}{2}$ such that $|N|\cdot\hat{d} > d$ and solve this problem: $\min_{0\le q_{i} < \hat{d}} \sum_{i}D_{i}(q_{i})$ subject to  $\sum_{i}q_{i} = d$.
Denote optimal value of this problem as $D^{\ast}_{\hat{d}}$. For each $i$, find ${\varepsilon}_{i}$ such that $D_{i}(q_{i}) \geq D^{\ast}_{\hat{d}}$ for all $q_{i} \in \left[\frac{d}{2} - {\varepsilon}_{i}, \frac{d}{2}\right)$. Such ${\varepsilon}_{i}$ always exists because $D_{i}(q_{i})$ is a strictly increasing function and $\lim_{q_{i}\to\frac{d}{2}} D_{i}(q_{i}) = \infty$. Therefore, we confer that the optimization problem in Theorem 3 is equivalent to this problem: $\min_{0\leq q_{i} \leq \frac{d}{2}-\varepsilon_{i}} \sum_{i}D_{i}(q_{i})$ subject to  $\sum_{i}q_{i} = d$, which has a unique solution. Therefore, the optimal solution always exists and the uniqueness follows from strict convexity of $D_{i}(q_{i})$. \\
Now we first note that 
\begin{equation} 
\label{eq8}
D^{'}_{i}(q_{i}) = \left(1 + \frac{q_{i}}{d - 2q_{i}}\right)C^{'}_{i}(q_{i})
\end{equation}
which is positive, strictly increasing function in $q_{i} \in \left[0, \frac{d}{2}\right)$. So, $D_{i}(q_{i})$ is strictly increasing and strictly convex function in $\left[0, \frac{d}{2}\right)$ because $D_{i}(q_{i}) = \int_{0}^{q_{i}} D^{'}_{i}(x_{i}) dx_{i} \geq C^{'}_{i}(0)\int_{0}^{q_{i}}\left(1 + \frac{x_{i}}{d} - 2x_{i}\right) dx_{i} = C^{'}_{i}(0)\int_{0}^{q_{i}}(\frac{1}{2}+\frac{d}{2d}-2x_{i})dx_{i} =  C^{'}_{i}(0)\int_{0}^{q_{i}}\left(\frac{1}{2q_{i}} - \frac{d}{4log(d-2x_{i})}\right) dx_{i}$. Thus, $\lim_{q_{i}\to \frac{d}{2}} D_{i}(q_{i}) = \infty$. Therefore, the optimization problem in the theorem is strictly convex problem and has unique optimal solution, and after a bit of mathematical manipulation, we get the unique solution $q^{\ast}$ determined by
%\begin{equation} 
%\label{eq9}
%\left(p^{\ast} -\left(1+\frac{q^{\ast}_{i}}{d - 2q^{\ast}_{i}}\right)C^{'}_{i}(q^{\ast}_{i})\right)\left(q_{i} - q^{\ast}_{i}\right) \leq 0, \forall q_{i}
%\end{equation}

%\begin{equation} 
%\label{eq10}
%\sum_{i}q^{\ast}_{i} = d
%\end{equation}

%\begin{equation} 
%\label{eq11}
%p^{\ast} > 0
%\end{equation}

%Also, note that we have
%\begin{equation} 
%\label{eq12}
 %\left(\frac{d}{B^{\ast}_{-i} + b^{\ast}_{i}} - \frac{B^{\ast}_{-i}}{B^{\ast}_{-i} - b^{\ast}_{i}}C^{'}_{i}\left(\frac{db^{\ast}_{i}}{B^{\ast}_{-i} + b^{\ast}_{i}}\right)\right)\left(b_{i} - b^{\ast}_{i} \right) \leq 0, \forall b_{i}
%\end{equation}
\begin{subequations}
\label{eq22}

\begin{equation}
\label{eq22a}
\left(p^{\ast} -\left(1+\frac{q^{\ast}_{i}}{d - 2q^{\ast}_{i}}\right)C^{'}_{i}(q^{\ast}_{i})\right)\left(q_{i} - q^{\ast}_{i}\right) \leq 0, \forall q_{i} 
\end{equation}

\begin{equation}
\label{eq22b}
\sum_{i}q^{\ast}_{i} = d 
\end{equation}

\begin{equation}
\label{eq22c}
p^{\ast} > 0 
\end{equation}
% \vspace{-2 mm}
\begin{equation}
%%%%%%---------?????? can be more smaller with:
% \small
% \footnotesize
% \scriptsize
% \tiny
%%%%%%---------??????
\begin{array}{rr}
\left(\frac{d}{B^{\ast}_{-i} + b^{\ast}_{i}} - \frac{B^{\ast}_{-i}}{B^{\ast}_{-i} - b^{\ast}_{i}}C^{'}_{i}\left(\frac{db^{\ast}_{i}}{B^{\ast}_{-i} + b^{\ast}_{i}}\right)\right)\left(b_{i} - b^{\ast}_{i} \right) \leq 0, \forall b_{i} \label{eq22d}
\end{array}
\end{equation}
\end{subequations}

Recall that the the Nash equilibrium value of $p^{\ast} = \frac{d}{\sum_{i}b^{\ast}_{i}}$ and the corresponding Nash equilibrium allocation $q^{\ast}_{i} = b^{\ast}_{i}p^{\ast}$. We can write (\ref{eq22d}) as $\left(p^{\ast} - \left(\frac{q^{\ast}_{i}}{d - 2q^{\ast}_{i}}\right)C^{\ast}_{i}(q^{\ast}_{i})\right)\left(b_{i}p^{\ast} - q^{\ast}_{i}\right) \leq 0$. Note that at the Nash equilibrium, $p^{\ast} > 0$ since $\sum_{i}b^{\ast}_{i} > 0$ by lemma 1. Thus the Nash equilibrium of the game satisfies (\ref{eq22a}) - (\ref{eq22c}), and solves the optimization problem in the theorem. The existence and uniqueness of the Nash equilibrium is a result of the existence and uniqueness of the optimal solution of the optimization problem.  $\blacksquare$
%\subsection{Proof of Theorem 4.}

\noindent\textbf{Proof of Theorem 4:} Note that $D^{'}_{i}(q_{i})$ is a strictly increasing function of $q_{i}$ and $D^{'}_{i}(0) = C^{'}_{i}(0)$. The proof follows the same argument as in Theorem 2. $\blacksquare$

%\subsection{Proof of Corollary 3.}

\noindent\textbf{Proof of Corollary 3:} From Theorem 4, we know that $\forall i \in \bar{N}$, $p^{\ast} = D^{'}_{i}(q^{\ast}_{i})$. Notice that $D_{i}(\cdot)$ is a strictly convex function. Thus, $D_{i}(q^{\ast}_{i}) - D_{i}(0) < D^{'}_{i}(q^{\ast}_{i})q^{\ast}_{i}$. Because $D_{i}(0) = 0$, $D_{i}(q) > C_{i}(q)$, we have $C_{i}(q^{\ast}_{i}) < p^{\ast}q^{\ast}_{i}$. $\blacksquare$

\noindent\textbf{Proof of Theorem 5:} Notice that $D^{'}_{i}(q_{i})$ and $C^{'}_{i}(q_{i})$ are both strictly increasing function and $D^{'}_{i}(q_{i}) \geq C^{'}_{i}(q_{i})$ for any $q_{i} \in \left[0, \frac{d}{2}\right)$. For any $i \in N$, $(D^{'}_{i})^{-1}(\bar{p}) \leq C^{-1}_{i}(\bar{p})$. Suppose $p^{\ast} < \bar{p}$. Because $C^{0}_{n^{\ast}} \leq p^{\ast} \leq C^{0}_{n^{\ast}+1}$, $C^{0}_{\bar{n}} \leq \bar{p} \leq C^{0}_{\bar{n}+1}$, and $C^{0}_{1} \leq C^{0}_{2} \leq ....... \leq C^{0}_{n}$, we have
 $n^{\ast} \leq \bar{n}$. Therefore, $\sum_{i}^{n^{\ast}}(D^{'}_{i})^{-1}(p^{\ast}) < \sum_{i}^{n^{\ast}}(D^{'}_{i})^{-1}(\bar{p}) \leq \sum_{i}^{n^{\ast}}(C^{'}_{i})^{-1}(\bar{p}) \leq \sum_{i}^{\bar{n}}(C^{'}_{i})^{-1}(\bar{p}) = d$, which contradicts that $\sum_{i}^{n^{\ast}}(D^{'}_{i})^{-1}(p^{\ast}) = d$. Thus, $p^{\ast} \leq \bar{p}$. Therefore, $\bar{n} \leq n^{\ast}$, implying $\bar{N} \subset N^{\ast}$. If $n^{\ast} < n$, then $p^{\ast} \leq D^{'}_{n^{\ast}+1}(0) \leq D^{'}_{n^{\ast}+1}\left(\frac{d}{n}\right) = \frac{n-1}{n-2}C^{'}_{n^{\ast}+1}\left(\frac{d}{n}\right) \leq \frac{n-1}{n} - 2M$. If $n^{\ast} = n$, there exists one $DH_{j}$ such that $0 < q^{\ast}_{j} \leq \frac{d}{n}$. Thus, $p^{\ast} = D^{'}_{j}(q^{\ast}_{i}) \leq D_{j}\left(\frac{d}{n}\right) \leq \frac{n-1}{n} - 2M$. In summary, 
\begin{equation} 
\label{eq13}
p^{\ast} \leq \frac{n-1}{n-2}M
\end{equation}

On the other side, there exists at least one $DH_{j}$ such that $C^{'}_{j}(\bar{q}_{i}) = \bar{p}$ and $\bar{q}_{i} \geq \frac{d}{n}$. Thus, 

\begin{equation} 
\label{eq14}
\bar{p} \geq C^{'}_{j}\left(\frac{d}{n}\right) \geq m
\end{equation}

Combing (\ref{eq13}) and (\ref{eq14}) gives  $p^{\ast} \leq \frac{n-1}{n-2}\frac{M}{m}\bar{p}$ . Lastly, $\bar{C} \leq C^{\ast}$ comes from the fact that $(\bar{q}_{i})_{i \in N}$ is an optimal solution of optimization problem in Theorem 1. If $\bar{q}_{max} < \frac{d}{2}$, then $\sum_{i}D_{i}(q^{\ast}_{i}) \leq \sum_{i}D_{i}(\bar{q}_{i})$ since $(q^{\ast})_{i \in N}$ is an optimal solution of optimization problem in Theorem 3. It is straightforward to check that $D_{i}(\bar{q}_{i}) \leq \left(1+\frac{\bar{q}_{i}}{d}-2\bar{q}_{i}\right)C_{i}(q^{\ast}_{i})$. Thus, $\sum_{i}D_{i}(q^{\ast}_{i}) \leq \left(1+\frac{\bar{q}_{max}}{d - 2\bar{q}_{max}}\right)\bar{C}$. On the other hand for any $q_{i} < \frac{d}{2}$, $D_{i}(q_{i}) = \left(1 + \frac{q_{i}}{d - 2q_{i}}\right)C_{i}(q_{i}) - \int_{0}^{q_{i}}\frac{d}{(d - 2x_{i})^2}C_{i}(x_{i})dx_{i} \geq \left(1 + \frac{q_{i}}{d - 2q_{i}}\right)C_{i}(q_{i}) - C_{i}(q_{i})\int_{0}^{q_{i}}\frac{d}{(d - 2x_{i})^2}dx_{i} \\ \geq  \left(1 + \frac{q_{i}}{d - 2q_{i}}\right)C_{i}(q_{i}) - C_{i}(q_{i})\frac{q_{i}}{d - 2q_{i}} \geq C_{i}(q_{i})$. \\Thus, $C^{\ast} = \sum_{i}C_{i}(q^{\ast}_{i}) \leq \sum_{i}D_{i}(q^{\ast}_{i}) \leq \left(1 + \frac{\bar{q}_{max}}{d - 2\bar{q}_{max}}\right)\bar{C}$. $\blacksquare$

\noindent\textbf{Proof of Theorem 6:} The proof of this theorem is dealt in various steps, the first step recognizing that proof directly follows from Theorem 1 in \cite{johari2009efficiency} due to the similarity in structure. 

\emph{Steps 2 of Proof of Theorem 6: A user's payoff is concave if he is price taking.} The condition that a uniform market-clearing price must exist implies that for any fixed $\theta > 0$, the range of $D(\mu , \theta)$ must contain $(0,\infty)$ as $\mu$ varies in $(0,\infty)$.
Now suppose that for fixed $\theta > 0$, there exist $\mu_1, \mu_2 >0$ with  $\mu_1 \neq \mu_2$ such that $D(\mu_1 , \theta)=D(\mu_2 , \theta)=d$, where $d>0$. Let $C=2d$ and let $R=2$. Then for $\pmb{\theta} =(\theta, \theta)$, there cannot exist a unique market-clearing price $p_D(\pmb{\theta})$; so we conclude that $D(\cdot, \theta)$ is monotonic, and strictly monotonic in the region where it is nonzero.

Let $I\subset (0,\infty)$ be the set of $\theta > 0$ such that $D(\mu , 0)$ is monotonically nondecreasing in $\mu$. From the preceding paragraph, we conclude that if $\theta\in (0,\infty) \backslash I$, then $D(\mu , \theta)$ is necessarily monotonically nonincreasing in $\mu$. Further, if $\theta \in I$, then $D(\mu , \theta) \rightarrow \infty$ as $\mu \rightarrow \infty$, and $D(\mu , \theta) \rightarrow 0$ as $\mu \rightarrow 0$; on the other hand, if   $\theta \in(0,\infty) \backslash I$, then $D(\mu , \theta) \rightarrow 0$ as $\mu \rightarrow \infty$, and $D(\mu , \theta) \rightarrow \infty$ as $\mu \rightarrow 0$.

Suppose $I\neq (0,\infty)$ and $I\neq \emptyset$; then choose $ \theta \in \partial I$, the boundary of $I$. Choose a sequence $\theta_n \in I$ such that $ \theta_n \rightarrow \theta$; and choose another sequence $\hat{\theta}_n\in (0,\infty) \backslash I$ such that $\hat{\theta}_n \rightarrow \theta$. Fix $\mu_1, \mu_2 $ with $0<\mu_1<\mu_2$, such that $D(\mu_1 , \theta)>0$ and $D(\mu_2 , \theta)>0$. Then we have $D(\mu_1 , \theta_n) \le D(\mu_2 , \theta_n)$, and  $D(\mu_1 , 
\hat{\theta}_n) \ge D(\mu_2 , \hat{\theta}_n)$. Taking limits as $n \rightarrow  \infty$, we get  $D(\mu_1 , \theta) \le D(\mu_2 , \theta)$, and  $D(\mu_1 , \theta) \ge D(\mu_2 , \theta)$, so that  $D(\mu_1 , \theta) = D(\mu_2 , \theta)$. But this is not possible, since $D(\cdot,\theta)$ must be strictly monotonic in the region where it is nonzero. Thus $I=(0,\infty)$ or $I=\emptyset$.

We will use Step 1 to show $D(\mu, \theta)$ is concave in $\theta\ge 0$ for fixed $\mu>0
$. Since $D(\mu, \theta)$ is continuous, it suffices to show that $D(\mu, \theta)$ is concave for $\theta >0$. Suppose not; fix $\theta >0, \overline{\theta}>0$, and $\delta\in(0,1)$ such that: 
\begin{equation}
    D(\mu, \delta\theta +(1-\delta)\overline{\theta})<\delta D(\mu, \theta)+(1-\delta)D(\mu, \overline{\theta})
    \label{EC.1}
    \tag{EC.1}
\end{equation}
Note this implies in particular that either $D(\mu, \theta)>0$ or $D(\mu, \overline{\theta})>0$. We assume without loss of generality  that $D(\mu, \theta)>0$. Let $C^R=RD(\mu, \theta)$, and let $\pmb{\theta}^R=(\theta,\cdots ,\theta)\in (\mathbb{R}^+)^R$. To emphasize the dependence of the market-clearing price on the capacity, we will let $p_D(\overline{\pmb{\theta}};C)$ denote the market-clearing price when the composite strategy vector is $\overline{\pmb{\theta}}$ and the capacity is C. We will show that for any $\theta '>0$, if $\mu^R=p_D(\pmb{\theta}^{R-1},\theta';C^R)$, then $\mu^R\rightarrow\mu $ as $R\rightarrow \infty$. First note that by definition, we have $D(\mu^R,\theta')+(R-1)D(\mu^R,\theta)=RD(\mu,\theta)$; or, rewriting, we have:
\begin{equation}
\frac{1}{R}D(\mu^R,\theta')+\left(1-\frac{1}{R}\right)D(\mu^R,\theta)=D(\mu,\theta)
\tag{EC.2}
\label{EC.2}
\end{equation}
Now note that as $R\rightarrow\infty$, the right hand side remains constant. Suppose that $\mu^R\rightarrow\infty$. Since $I=(0,\infty)$ or $I=\emptyset$, either $D(\mu^R,\theta'),D(\mu^R,\theta)\rightarrow 0$, or $D(\mu^R,\theta'),D(\mu^R,\theta)\rightarrow \infty$; in either case, the equality 
(\ref{EC.2}) is violated for large R. A similar conclusion holds if $\mu^R\rightarrow 0$ as $R\rightarrow\infty$. Thus we do not have $\mu^R\rightarrow 0$ or $\mu^R\rightarrow \infty$as $R\rightarrow\infty$. Choose a convergent subsequence, such that $\mu^R_k\rightarrow\hat{\mu}$, where $\hat{\mu}\in(0,\infty)$. From 
(\ref{EC.2}), we mush have $D(\hat{\mu}, \theta)=D(\mu, \theta)$. But as established above, since $D(\cdot, \theta)$ is strictly monotonic in the region where it is nonzero, this is only possible if $\hat{\mu}=\mu$. We conclude that the following three limits hold:
$$\lim_{R\rightarrow \infty} p_D(\pmb{\theta}^R; C^R)=\mu;$$
$$\lim_{R\rightarrow \infty} p_D(\pmb{\theta}^{R-1},\overline{\theta}; C^R)=\mu;$$
$$\lim_{R\rightarrow \infty} p_D(\pmb{\theta}^{R-1},\delta\theta +(1-\delta)\overline{\theta}); C^R)=\mu;.$$

The remainder of the proof is straightforward. From 
(\ref{EC.1}), for $R$ sufficiently large, we must have:
$$D(p_D(\pmb{\theta}^{R-1},\delta\theta +(1-\delta)\overline{\theta}); C^R),\delta\theta +(1-\delta)\overline{\theta})$$
$$<
\delta D(p_D(\pmb{\theta}^{R}; C^R),\theta) +(1-\delta)D(p_D(\pmb{\theta}^{R-1},\overline{\theta}; C^R),\overline{\theta}).$$
This violates the conclusion of Step 1, so we conclude $D(\mu, \theta)$ is concave in $\theta\ge 0$ give $\mu>0$. A similar argument shows that $\mu D(\mu, \theta)$ is convex in $\theta$, by using the fact that $p_D(\pmb{\theta})D(p_D(\pmb{\theta}), \theta_r)$ must be convex in $\theta_r$ for nonzero $\pmb{\theta}$. Combining these results yields the desired conclusion.

\emph{Step 5, Proof of Theorem 6: B is an invertible, differentiable, strictly increasing, and concave function on $(0,\infty)$.} Note from 
(10) %!?
that:
\begin{equation}
    B(p_D(\pmb{\theta}))=\frac{\sum_{r=1}^{R}\theta_r}{C}.
    \tag{EC.3}
    \label{EC.3}
\end{equation}
We immediately see that  B must be invertible on $(0,\infty)$; it is clearly onto, as the right hand side of (\ref{EC.3}) can  take  any  value  in  $(0,\infty)$. Furthermore, if  $B(p_1)=B(p_2)=\gamma$
for some prices $p_1,p_2>0$, then choosing $\pmb{\theta}$ such that $\sum_{r=1}^{R}\theta_r/C=\gamma$, we find that $p_D(\pmb{\theta})$ is not uniquely defined. Thus B is one-to-one as well, and hence invertible. Finally, note that since D is differentiable, B must be differentiable as well. 
We let $\Phi$ denote  the  differentiable  inverse  of  B. We  will  show  that  $\Phi$ is strictly increasing and convex. We first note that for nonzero $\pmb{\theta}$ we have:
$$p_D(\pmb{\theta})=\Phi\left(\frac{\sum_{r=1}^{R}\theta_r}{C}\right).$$
Let 
\begin{equation}
%%%%%%---------?????? can be more smaller with:
% \small
% \footnotesize
% \scriptsize
% \tiny
%%%%%%---------??????
\begin{array}{rr}
   w_r(\pmb{\theta})= p_D(\pmb{\theta}) D(p_D(\pmb{\theta}) ,\theta_r)
=\Phi\left(\frac{\sum_{s=1}^{R}\theta_s}{C}\right)
\left(\frac{\theta_r}{\sum_{s=1}^{R}\theta_s} C\right)\\
\tag{EC.4}
\label{EC.4}
\end{array}
\end{equation}
By Step 1, $w_r(\pmb{\theta})$ is  convex  in $\theta_r>0$. By considering strategy vectors 
$\pmb{\theta}$ for which $\pmb{\theta}_{-r}=0$, it follows that $\Phi$ is convex.

It remains to be shown that $\Phi$ is strictly increasing. Since $\Phi$  is invertible, it must be monotonic; and thus $\Phi$  is either strictly increasing or strictly decreasing. To simplify the argument, we assume that $\Phi$ is twice differentiable. We twice differencetiate  $w_r(\pmb{\theta})$, given in 
(\ref{EC.4}). Letting $\mu=\sum_{s=1}^{R}\theta_s/C$, we have for nonzero $\pmb{\theta}$:
\begin{equation}
    \frac{\partial^2 w_r}{\partial\theta_r^2}(\pmb{\theta})=\Phi^{''}(\mu)
\frac{\theta_r}{C^2\mu}+\frac{2\sum_{s\neq r}\theta_s}{C^2\mu^3}(\mu \Phi'(\mu)-\Phi(\mu)).
\label{EC.5}
\tag{EC.5}
\end{equation}
Consider some nonzero $\pmb{\theta}_{-r}$, and take the limit as $\theta_r\rightarrow 0$. The limit of the left-hand side in (\ref{EC.5}) is nonnegative, by the convexity of $w_r(\pmb{\theta})$ in $\theta_r>0$. The limit of the first term in the right-hand side of 
(\ref{EC.5}) is zero. 
Since $\Phi(\mu)>0$, it follows that $\Phi'(\mu)>0$, so that $\Phi$ is strictly increasing. This establishes the desired facts regarding B.

\emph{Steps 6, Proof of Theorem 6: Let (C,R,\textbf{U}) be a utility system. A vector $\pmb{\theta}\ge 0$ is a Nash equilibrium if and only if at least two components of $\pmb{\theta}$ are nonzero, and there exists a nonzero vector $\pmb{d}\ge 0$ and a scalar $\mu>0$ such that $\theta_r=\mu d_r$ for all r, $\sum_{r=1}^R d_r=C$, and the following conditions hold:
$$
%%%%%%---------?????? can be more smaller with:
% \small
% \footnotesize
% \scriptsize
% \tiny
%%%%%%---------??????
\footnotesize{ 
\begin{aligned}
    U'_r(d_r)\left(1-\frac{d_r}{C}\right)
=\Phi(\mu)
\left(1-\frac{d_r}{C}\right)
+\mu\Phi'(\mu)
\left(\frac{d_r}{C}\right),& \textit{  if } d_r>0;
\\
U'_r(0)\le\Phi(\mu), &\textit{  if } d_r=0.
\end{aligned}
}
$$
In this case $d_r=D(p_D(\pmb{\theta}),\theta_r))$, $\mu = \sum_{r=1}^R \theta_r/C$, and $\Phi(\mu)=p_D(\pmb{\theta})$.} Suppose that $\pmb{\theta}$ is a Nash equilibrium. Since $Q_r(\theta_r;\pmb{\theta}_{-r})=-\infty$ if $\pmb{\theta}=0$, (from (7)), we must have $\pmb{\theta}\neq 0$. Suppose then that only one component of $\pmb{\theta}$ is nonzero; say $\theta_r>0$, and $\pmb{\theta}_{-r}=0$. Then the payoff to user r is:
$$U_r(C)-\Phi\left(\frac{\theta_r}{C}\right)C$$
But now observe that by infinitesimally reducing $\theta_r$, user $r$ can strictly improve his payoff (since $\Phi$ is strictly increasing). Thus $\pmb{\theta}$ could not have been a Nash equilibrium; we conclude that at least two components of $\pmb{\theta}$ are nonzero. In this case, from (7), and the expressions in (11) and (\ref{EC.4}), the payoff $Q_r(\overline{\theta}_r;\pmb{\theta}_{-r})$ to  user  $r$ is differentiable. When two components of $\pmb{\theta}$ are nonzero, we may write the payoff $Q_r$ to user $r$ as follows, using (11) and (\ref{EC.4}):
$$
%%%%%%---------?????? 
\begin{array}{ll}
Q_r(\theta_r;\pmb{\theta}_{-r})
&=U_r\left(\frac{\theta_r}{\sum_{s=1}^R \theta_s}C\right)\\
&-\Phi\left(\frac{\sum_{s=1}^R \theta_s}{C}\right)\left(\frac{\theta_r}{\sum_{s=1}^R \theta_s}C\right).
\end{array}
$$
Differentiating  the  previous expression with respect to $\theta_r$, we conclude that if $\pmb{\theta} $ is a Nash equilibrium then the following optimality conditions hold for each $r$:
\begin{equation}
    F_r(\pmb{\theta})=0 \textit{ if } \theta_r>0;
    \tag{EC.6}
    \label{EC.6}
\end{equation}
\begin{equation}
    F_r(\pmb{\theta})\le0 \textit{ if }  \theta_r=0,
    \label{EC.7}
\end{equation}

where 
$$
\begin{array}{ll}
F_r(\pmb{\theta}) &= U'_r\left(\frac{\theta_r}{\sum_{s=1}^R \theta_s}C\right)
\left(\frac{C}{\sum_{s=1}^R \theta_s}-\frac{\theta_r C}{(\sum_{s=1}^R \theta_s)^2}\right) \\
&-
\Phi'\left(\frac{\sum_{s=1}^R \theta_s}{C}\right)\left(\frac{\theta_r}{\sum_{s=1}^R \theta_s}\right)\\
 &-
    \Phi\left(\frac{\sum_{s=1}^R \theta_s}{C}\right)\left(\frac{C}{\sum_{s=1}^R \theta_s}-\frac{\theta_r C}{(\sum_{s=1}^R \theta_s)^2}\right)
\end{array}
$$
These conditions are equivalent to (14)-(15), if we make the substitutions $\mu=\sum_{s=1}^R \theta_s/C$, and $d_r=D(p_D(\pmb{\theta}),\theta_r)$. Furthermore, in this case we have $\pmb{d}\ge 0
, \mu >0, \theta_r=\mu d_r, \sum_{r=1}^R d_r=C$,  and $p_D(\pmb{\theta})=\Phi(\mu)$.

On the other hand, suppose that we have found $\pmb{\theta,d}$ and $\mu$ such that the conditions of Step 6 are satisfied. In this case we simply reverse the argument above; since $Q_r(\overline{\theta}_r;\pmb{\theta}_{-r})$ is concave in $\overline{\theta_r}$ (Condition 2 in Definition 4), if at least two components of $\pmb{\theta} $ are nonzero then the conditions (\ref{EC.6})-(\ref{EC.7}) are necessary and sufficient for $\pmb{\theta} $ to be a  Nash equilibrium. Furthermore, if $\pmb{d}\ge 0,  \mu >0, \theta_r=\mu d_r$, and $ \sum_{r=1}^R d_r=C$,  then it follows that $\mu=\sum_{s=1}^R \theta_s/C$, $\Phi(\mu)=p_D(\pmb{\theta})$, and 
$d_r=D(p_D(\pmb{\theta}),\theta_r)$. Thus the conditions  (\ref{EC.6})-(\ref{EC.7}) become equivalent to (14)-(15), as required.

\emph{Steps 7, Proof of Theorem 6: Let (C,R,\textbf{U}) be a utility system. Then there exists a unique   Nash equilibrium.} Our approach will be to demonstrate existence of a  Nash equilibrium by finding a solution $\mu >0$ and $\pmb{d}\ge 0$ to (14)-(15),  such that $\sum_{r=1}^R d_r=C$. If we find such a solution, then at least two components of $\pmb{d}$ must be nonzero; otherwise, (14) cannot hole for the user $r$ with $d_r=C$. If we define $\pmb{\theta}=\mu d $, then $\mu=\sum_{s=1}^R \theta_s/C$, so $p_D(\pmb{\theta})=\Phi(\mu)$; and from (11), we have $d_r=D(p_D(\pmb{\theta}),\theta_r)$. Thus if $\mu >0$ and $\pmb{d}\ge 0$ satisfy (14)-(15), then $\pmb{\theta}=\mu d $ is a   Nash equilibrium by Steps 6. Consequently, it suffices to find a solution $\mu >0$ and $\pmb{d}\ge 0 $ to (14)-(15).

We first show that for a fixed value of $\mu >0$ , the equality in (14) has at most one solution $d_r$. To see this, rewrite (14) as:
$$U'_r(d_r)\left(1-\frac{d_r}{C}\right)
-(\mu\Phi'(\mu)-\Phi(\mu))\left(\frac{d_r}{C}\right)=\Phi(\mu).
$$
Since $\Phi$ is convex and strictly increasing with $\Phi(\mu)\rightarrow0$ as $\mu \rightarrow 0$, we have $\mu\Phi'(\mu)-\Phi(\mu) \ge 0$. Thus the left hand side is strictly decreasing in $d_r$ (since $U_r$ is strictly increasing and concave), from $U'_r(0)$ at $d_r=0$ to 
$\mu\Phi'(\mu)-\Phi(\mu) \le 0$ when $d_r=C$. This implies a unique solution $d_r\in [0,C] $ exists for the equality in (14) as long as $U'_r(0) \ge \Phi(\mu)$; we denote this solution $d_r(\mu)$. If $ \Phi(\mu)>U'_r(0) $, then we let $d_r(\mu)=0$. Observe that as $\mu \rightarrow0$, we must have $d_r(\mu)\rightarrow C$, since otherwise we can show that (14) fails to hold for sufficiently small $\mu$.

Next we show that $d_r(\mu)$ is continuous. Since we defined $d_r(\mu)=0$ if $ \Phi(\mu)>U'_r(0) $, and $d_r(\mu)=0$ if $\Phi(\mu)=U'_r(0) $ from (14), it suffices to show that $d_r(\mu)$ is continuous for $\mu$ such that  $\Phi(\mu)\le U'_r(0) $. But in this case continuity of $d_r$ can be shown using (14), together with the fact that $U'_r, \Phi$ and $\Phi'$ are all continuous (the latter because  $\Phi$ is concave and differentiable, and hence continuously differentiable). Indeed, suppose that $\mu_n \rightarrow \mu $ where $\Phi(\mu)\le U'_r(0) $, and assume without loss of generality that $d_r(\mu_n)\rightarrow d_r$ (since  $d_r(\mu_n)$ takes values in the compact set [0,C]). Then since $\mu_n$ and $d_r(\mu_n)$ satisfy the equality in (14) for sufficiently large n, by taking limits we see that $\mu$ and $d_r$ satisfy the equality in (14) as well. Thus we must have $d_r=d_r(\mu)$, so we conclude $d_r(\mu)$ is continuous.

We now show that $d_r(\mu)$ is nonincreasing in $\mu$. To see this, choose $\mu_1,\mu_2>0$ such that $\mu_1<\mu_2$. Suppose that $d_r(\mu_1)<d_r(\mu_2)$. Then, in particular, $d_r(\mu_2)>0$, so (14) holds with equality for $d_r(\mu_2)$ and $\mu_2$. Now note that as we move from $d_r(\mu_2)$ to $d_r(\mu_1)$, the left hand side of (14) strictly increases (since $U_r$ is concave). On the other hand, since $\Phi$ is convex and strictly increases with $\Phi(\mu )\rightarrow 0$ as $\mu \rightarrow 0$, we have the inequalities 
$\mu_2 \Phi'(\mu_2)-\Phi(\mu_2) \ge \mu_1 \Phi'(\mu_1)-\Phi(\mu_1) \ge 0$. From this it follows that the right hand side of (14) strictly decreases as we move from $d_r(\mu_2)$ to $d_r(\mu_1)$ and  from $\mu_2$ to $\mu_1$. Thus  neither (14) nor (15) can hold at $d_r(\mu_1)$
and $\mu_1$; so we conclude that for all $r$, we must have $d_r(\mu_1)\ge d_r(\mu_2)$.

Thus for each $r$, $d_r(\mu)$ is a nonincreasing continuous function such that $d_r(\mu) \rightarrow C$ as $\mu \rightarrow 0$, and $d_r(\mu) \rightarrow 0$ as $\mu \rightarrow \infty$. We conclude there exists at least one $\mu>0$ such that $\sum_{r=1}^R d_r(\mu)=C$; and in this case $\pmb{d}(\mu)$ satisfies (14)-(15), so by the discussion at the beginning of this step, we know that $\pmb{\theta} = \mu \pmb{d}(\mu) $ is a Nash equilibrium.

Finally, we show that the Nash equilibrium is unique. Suppose that there exist two solutions $\pmb{d}^1\ge 0, \mu_1>0$, and $\pmb{d}^2\ge 0, \mu_2>0$ to (14)-(15), such that $\sum_{r=1}^R d_r^i=C$ for $i=1,2.$ Of course, we must have $\pmb{d}^i=\pmb{d}(\mu_i), i = 1,2.$ We assume without loss of generality that $\mu_1\le \mu_2$; our goal is to show that $\mu_1=\mu_2$ . Since $d_r(\cdot)$ is nonincreasing, we know $d_r(\mu_1)\ge d_r(\mu_2)$ for all $r$. Since 
$\sum_{r=1}^R d_r^i=C$ for $i=1,2,$ we conclude that  $d_r(\mu_1)= d_r(\mu_2)$ for every $r$. Let $r$ be such that $d_r(\mu_1)= d_r(\mu_2)>0$. Observe that $\Phi(\mu)$ and $\mu\Phi'(\mu)$ are both strictly increasing in $\mu>0$, since $\Phi$ is strictly increasing and convex. Thus for fixed $d_r>0$, the equality in (14) has a unique solution $\mu$, so $d_r(\mu_1)= d_r(\mu_2)>0$ implies $\mu_1=\mu_2$. Thus (14)-(15) have a unique solution $\pmb{d}\ge 0, \mu>0$, such that  $\sum_{r=1}^R d_r=C$. From Step 6, this ensures the Nash equilibrium $\pmb{\theta}=\mu \pmb{d}$ is unique as well.
Thus, combining steps 1 to 7, we prove Theorem 6. $\blacksquare$

% \bibliographystyle{unsrt}
%\bibliographystyle{ACM-Reference-Format}
% \bibliography{scibib1,scibib2}

% johari2009efficiency

%\bibliographystyle{ACM-Reference-Format}
%\bibliography{appendixPER}

\bibliographystyle{unsrt}
\bibliography{arxiv}

\end{document}